\documentclass[pdflatex,sn-basic,Numbered]{sn-jnl}

\usepackage{graphicx}%
\usepackage{multirow}%
\usepackage{amsmath,amssymb,amsfonts}%
\usepackage{amsthm}%
\usepackage[title]{appendix}%
\usepackage{xcolor}%
\usepackage[table]{xcolor}
\usepackage{textcomp}%
\usepackage{manyfoot}%
\usepackage{booktabs}%
\usepackage{algorithm}%
\usepackage{algorithmicx}%
\usepackage{algpseudocode}%
\usepackage{listings}%
\usepackage{longtable}%
\usepackage{array}%
\usepackage[T1]{fontenc}%
\usepackage[utf8]{inputenc}%
\usepackage{tabularx}
\usepackage{makecell}
\usepackage{float}

\definecolor{idcell}{RGB}{220,235,255}
\definecolor{lencell}{RGB}{255,242,204}
\definecolor{payloadcell}{RGB}{226,239,218}
\definecolor{checksumcell}{RGB}{252,228,214}
\definecolor{othercell}{RGB}{242,242,242}

\begin{document}

\title[Unified Message Model]{Unified Message Model for Heterogeneous Serial Data Exchange Protocols}

\author*[1]{\fnm{Viktor} \sur{Sinitsyn}}\email{viktor.sinitsyn@tum.de}
\author[1]{\fnm{Florian} \sur{Holzapfel}}\email{florian.holzapfel@tum.de}

\affil*[1]{\orgname{Institute of Flight System Dynamics, Technical University of Munich}, \orgaddress{\city{Munich}, \country{Germany}}}

\abstract{Modern embedded systems are becoming increasingly complex and typically integrate numerous heterogeneous devices, such as controllers, sensors, actuators, and supporting subsystems. As a result, their development and integration involve a wide variety of serial communication protocols, ranging from standardized solutions to partially standardized and fully project-defined formats. Efficient development of such systems increasingly depends on automation toolchains, which in turn require a clear, unified, and machine-processable formal basis. 
This paper proposes a unified, protocol-agnostic message model for explicit and deterministic description of serial messages. The model is based on formal definition of data types, atomic message elements (containers), and complete message structure. In addition to the model itself, the paper introduces methods for practical work with it, including configurable message types for expressing structural constraints and supporting deterministic automation, as well as configurable user representations for engineering-oriented reading and editing. 
The proposed model and methods are demonstrated through implementation in an industrial tool environment. The results show that the approach can support machine-readable interface control document development, automated generation of transport-layer software, and practical engineering work with both standardized and weakly formalized serial protocols. Taken together, the proposed model, methods, and tool implementation provide a practical foundation for automation toolchains in heterogeneous serial communication development.}


\maketitle

\section{Introduction}

\subsection{Heterogeneous Serial Protocols in Embedded Systems}

Modern embedded electronic systems are becoming increasingly complex and increasingly distributed. Instead of a small number of tightly integrated components, they often consist of many interacting devices, sensors, controllers, and supporting subsystems \cite{007,032}. As a result, system integration and development are becoming more difficult as well \cite{048}.

Historically, such challenges were most typical of large aerospace systems. Today, similar development burden can also be observed in rapidly developing unmanned aerial vehicle (UAV) platforms, air taxis, and other highly automated embedded systems \cite{007}.

Integration activities of such systems typically require the use of diverse serial communication protocols. These range from strictly standardized solutions, such as ARINC 429 or CAN, to fully project-defined formats implemented over generic serial interfaces such as universal asynchronous receiver-transmitter (UART).

\subsection{Engineering Problem and Practical Motivation}

One of the most labor-intensive engineering problems in heterogeneous systems integration is interface development, which largely consists of repeated transformation of the same interface information from one representation into another. Source interface descriptions received from equipment vendors in vendor-specific formats must be reworked into interface control documents (ICDs) and then further transformed into transport-layer software (TL SW) that encodes engineering values into protocol byte streams and decodes protocol byte streams back into engineering values used by applications. In practice, this transformation is performed repeatedly for redundant channels of onboard equipment and supporting environments such as hardware-in-the-loop test rigs, flight test instrumentation (FTI), and laboratory analyzers \cite{037, 041, 044, 048, 052}.

Repeated manual processing of the same ICD data, whether through conversion of human-readable document formats or through manual development of TL SW source code, results in substantial engineering effort and multiple human-induced errors \cite{052, 058}. For this reason, efficient practical development relies on automation toolchains used for the generation of TL SW and, where needed, corresponding human-readable ICDs \cite{007, 037, 041, 048}. Similar tool-supported and model-based system engineering (MBSE) approaches have long been pursued in large aerospace companies and safety-critical development environments \cite{037, 038, 040, 049, 051}.

A necessary core element of such a toolchain is a formalized machine-readable ICD format that can serve as a common source for downstream engineering artifacts \cite{037, 041, 044, 057}. A classical way to define such an ICD is to represent it as a collection of protocol-specific parts such as CAN \cite{s2}, Ethernet \cite{s4, s5, s6}, or ARINC 429 \cite{s1}. However, this approach has several important limitations.

\begin{itemize}
\item It cannot be applied as is to protocols for which no accepted message standard exists. In such cases, a new protocol-specific part must be developed for each new protocol type.
\item It cannot adequately cover practical implementations that deviate from the standard in specific parts, such as CAN-based messages with little-endian payload encoding, without additional effort to adapt the toolchain.
\item It leads to unnecessary implementation overhead, because supporting tools must maintain separate processing branches for different protocol families and repeatedly adapt them to deviations from standards encountered in practice.
\item It provides no unified basis for automation across heterogeneous communication environments. In practice, different protocol families require different description mechanisms and tools, which substantially increases toolchain development and maintenance effort within a single project.
\end{itemize}
For this reason, machine-readable ICD and automated TL SW development would benefit from a unified, protocol-agnostic message model capable of describing arbitrary message formats within a single common framework.

\subsection{Existing Description Approaches and Their Limitations}

Several established approaches exist for describing communication interfaces and message formats. In the automotive domain, AUTOSAR \cite{s7} defines interface descriptions through ARXML tightly integrated with its architectural framework. For CAN-based systems, DBC \cite{s8} files are widely used to describe signal layouts within frames. Formal description languages such as ASN.1 \cite{s9} provide standardized mechanisms for defining message structures and encoding rules and are extensively applied in telecommunications and related domains.

Despite their usefulness, these approaches do not close the gap between standardized message specification mechanisms and the practical needs of heterogeneous embedded system integration. Existing solutions typically exhibit one or more of the following characteristics:

\begin{itemize}
\item tight coupling to a particular ecosystem or architectural framework, as in AUTOSAR-based interface descriptions;
\item addressing only a specific protocol family, as in CAN-oriented DBC definitions;
\item implicit reliance on assumptions inherited from a particular standard or modeling approach, for example specifying a message identifier value without explicitly defining how and where that identifier is encoded in the transmitted message;
\item assumptions of predefined encoding grammar optimized for telecommunications standards, as in ASN.1.
\end{itemize}
As a result, heterogeneous protocol sets within a single engineering project are still typically handled by combining multiple protocol-specific description approaches. In practice, embedded system developers frequently encounter a mixture of partially standardized and fully custom serial message formats that must be implemented across onboard software, redundant channels, and multiple supporting environments. Under such conditions, no commonly adopted protocol-agnostic message meta-model exists that:

\begin{itemize}
\item uniformly describes arbitrary serial message structures across heterogeneous protocols;
\item explicitly captures low-level encoding semantics required for implementation;
\item serves simultaneously as a machine-processable source for automated code generation and as a structured basis for system-level interface management.
\end{itemize}
\subsection{Special Case of Weakly Formalized Protocols}

This limitation becomes especially critical for weakly formalized protocols. For communication types such as ARINC 429, CAN, ARINC 825 \cite{s3}, MAVLink \cite{s15} or UDP/IP, at least a formalized frame and header structure exists and can serve as a reference for interface description. In contrast, UART-based and similar project-defined serial formats often have no commonly adopted formal message description approach at all. Message sizes, endianness, encoding rules, and addressing conventions are then defined descriptively in human-readable documents on a project-specific basis.

For such protocols, the absence of a unified description method makes systematic automation particularly difficult. In these cases, a unified message model is not merely a way to simplify existing practice, but a necessary basis for introducing a consistent and machine-processable description at all.

\subsection{Contribution of This Paper}

This paper addresses the problem described through three complementary contributions.

First, it introduces a unified, protocol-agnostic meta-model for describing serial message structures in an explicit and machine-processable form. The model is intended to support arbitrary message formats across heterogeneous communication environments.

Second, the paper introduces methods for working with this model in practice. These methods include mechanisms for defining structural constraints, supporting deterministic automation, and providing configurable engineering-oriented representations of the same underlying message description. In this way, the work addresses not only the problem of formal representation, but also the problem of efficient use of such a representation in real development workflows.

Third, the proposed concepts are applied in an industrial MBSE tool, providing a reference implementation of the model and methods.

Taken together, these contributions establish all three major elements of a model-based systems engineering (MBSE) approach: model, methods, and tool support.

\subsection{Structure of the Paper}

The remainder of the paper is organized as follows. Section 2 introduces the terminology used throughout the paper. Section 3 reviews the state of the art, including existing standards and related approaches to interface and message description. Section 4 presents the proposed unified message model. Section 5 introduces the methods for working with this model in practice, including mechanisms for structural constraints, automation, and configurable user representations. Section 6 presents the reference implementation and demonstrates its application in practical engineering workflows. Finally, Section 7 summarizes the results.

\section{Terminology}

This section introduces several terms used throughout the paper to describe the structure of serial communication messages.

{\small
\begin{longtable}{p{0.22\textwidth}p{0.72\textwidth}}
\caption{Terminology used in the paper}\label{tab:1}\\
\toprule
Term & Description \\
\midrule
\endfirsthead

\caption[]{Terminology used in the paper (continued)}\\
\toprule
Term & Description \\
\midrule
\endhead

\bottomrule
\endfoot

\bottomrule
\endlastfoot

Container & A container is a structural component of a message that encapsulates a logically related portion of the message data. Containers may represent message identifiers, payload elements, length fields, checksums, or other structural elements. \\
Container role & A container role denotes the functional purpose of a container within a message structure, such as payload, identifier, length, checksum, or other. \\
Data type & A data type defines the set of possible values of a message element and the basic rules of their encoded representation and interpretation. In the proposed model, data types serve as the basic encoding units referenced by message containers. \\
Decoding & Decoding is the process of extracting and interpreting encoded message contents in order to reconstruct engineering-level values. \\
Encoding & Encoding is the process of transforming engineering-level values into their encoded representation within a message according to the applicable encoding rules. \\
Encoding semantics & Encoding semantics denotes the set of properties that define how the value of a message element is encoded and interpreted, including aspects such as data type, bit length, signedness, sign encoding, byte order, bit order, resolution, and offset. \\
Embedded System & An Embedded system is a computer-based technical system designed to perform dedicated operational functions within a larger physical system and typically integrating computing devices, sensors, actuators, and communication interfaces. A representative example is onboard aircraft equipment, including UAV platforms. \\
Heterogeneous Serial Protocols & Heterogeneous serial protocols are serial communication protocols that are based on different underlying communication standards or project-specific conventions and therefore differ in frame structure, header organization, identifier representation, encoding semantics, and payload layout. In the context of this paper, the term emphasizes that a unified message model must be applicable across such structurally different protocols within one common framework. \\
Identifier & An identifier is a container role that is used to distinguish a message type or message instance within a communication channel. \\
Message & A message is a unit of data transmitted as a single communication frame or packet over a serial interface. A message consists of a sequence of structured data elements that collectively represent a specific data exchange instance between communicating systems. \\
Message model & A message model is a structured representation of a complete message, combining message-level properties with a set of containers that together define its full encoded form. \\
Message structure & Message structure describes the ordered arrangement of containers that together form a complete message. \\
Message type & A message type is a formal description of allowable characteristics and structural constraints for a class of messages within a given communication domain. \\
Payload & Payload refers to the portion of a message that carries application-level data intended for consumption by the receiving system. The term can also refer to the role of containers carrying such data. \\
Serial Protocols & Serial protocols are digital communication protocols in which information is transmitted as an ordered sequence of bits or bytes over a communication channel according to defined structural and encoding rules. \\
Transport-layer software & Transport-layer software is software responsible for encoding engineering values into protocol byte streams and decoding protocol byte streams back into engineering values used by applications. \\
User representation & A user representation is a configurable engineering-oriented view of the underlying formal message model, intended for convenient reading, editing, and practical use. \\

\end{longtable}
}

\section{State-of-the-Art}

This section reviews the state of the art relevant to the present work in three steps. Section~3.1 considers communication standards that define message transmission and, to different degrees, message structure. Section~3.2 reviews standards and formats intended for interface description and automation-oriented development toolchains. Section~3.3 then considers published work addressing system complexity, MBSE-based development, automation approaches, and machine-readable handling of interface and message data. This structure is intended to position the proposed model both with respect to existing standards and with respect to related engineering methods and tools.

\subsection{Serial Protocols Standards}

A number of communication standards already provide formalized definitions of message transmission and message structure. For the purposes of this paper, they are important not because they are insufficient in some absolute sense, but because together they illustrate the variety of communication formats that a unified message model should be able to describe within one common framework.

\subsubsection{ARINC 429}

ARINC 429 \cite{s1} is a highly formalized avionics communication standard. It defines electrical characteristics, timing, word structure, and protocol rules for 32-bit words transmitted over the bus. It therefore represents a communication format with a strongly constrained and explicit message structure. From the perspective of the present work, ARINC 429 is important as an example of a standard whose message definition is already highly formalized and can be represented in a unified way.

\subsubsection{CAN}

Classic CAN \cite{s2} standardizes frame structure, identifiers, control fields, data fields, and error handling, which makes it a strong example of a protocol family with a clearly defined message format. In the context of this paper, CAN is important not only as a standardized protocol in its own right, but also as a basis for widely used higher-level development ecosystems.

\subsubsection{ARINC 825}

ARINC 825 \cite{s3} is an airborne communication standard built on CAN. It uses the CAN extended frame format as its communication basis, while introducing additional aerospace-specific conventions and constraints. This makes ARINC 825 especially relevant to the present work: although it shares the same underlying frame basis with CAN EFF, its engineering interpretation and usage constraints differ. A unified message model should therefore be able to represent both the shared structural basis and the domain-specific differences built on top of it.

\subsubsection{Ethernet, IP, and UDP}

These standards \cite{s4, s5, s6} form a widely used communication stack. Ethernet provides link-layer framing and addressing, IP defines packet addressing and forwarding at the network layer, and UDP defines a lightweight transport-layer header and datagram mechanism. At the same time, this stack mainly standardizes communication headers and transmission-layer conventions, while the application-level payload structure remains outside its scope. For the present work, this stack is therefore important as an example of a communication environment in which lower communication layers are standardized, while the engineering content of transmitted messages still requires an additional explicit description model.

\subsubsection{UART}

Universal asynchronous receiver-transmitter differs from the standards above in that it defines only a serial transmission mechanism with byte framing and related low-level communication parameters. It does not define application-level message structure. In practice, message boundaries, field ordering, endianness, integrity checks, and payload interpretation are usually specified on a device or project-specific basis. For this reason, UART-based formats are one of the clearest examples of communication environments for which a unified transport-layer message model is especially valuable.

\subsubsection{MAVLink}

MAVLink \cite{s15} is a binary telemetry protocol used in unmanned and robotic systems to provide a standardized message structure over otherwise simple communication links like UART or UDP payload. Its role is not to define the underlying physical transport, but to organize application-level message exchange within a specific domain in a consistent and interoperable way. In the context of the present work, it is relevant as an example of a domain-specific attempt to standardize the structure of otherwise weakly formalized serial messages. 

\subsubsection{Summary}

Taken together, these standards demonstrate that communication environments may differ substantially in scope, level of formalization, and layer coverage. Some define message structure in detail, some define only headers or transport mechanisms, and some leave most message semantics to project-specific conventions. The objective of the proposed work is therefore not to replace these standards, but to provide a unified message model capable of describing all such cases within one explicit and machine-processable framework.

\subsection{Description Formats and Toolchain-Oriented Standards}

In addition to communication standards themselves, a number of established formats and frameworks are used to describe communication interfaces and support engineering automation.

\subsubsection{AUTOSAR}

AUTOSAR \cite{s7} is not a communication protocol standard, but a comprehensive framework and toolchain-oriented architecture for automotive electronic control unit (ECU) development. Official AUTOSAR documentation describes it as an architecture for automotive ECUs connected to networks such as CAN \cite{s2}, LIN \cite{s11}, FlexRay \cite{s12}, and Ethernet \cite{s4, s5, s6}. In this sense, AUTOSAR is a strong positive example: it provides exactly the kind of tool-supported development environment that is needed in complex projects. However, its applicability is tied to the AUTOSAR ecosystem, its architectural concepts, and its target domain. It is therefore not a protocol-agnostic basis for describing arbitrary serial message structures such as UART-based custom formats or avionics standards such as ARINC 429.

\subsubsection{DBC}

DBC \cite{s8} is a simple and widely used format for describing CAN communication. In practice, it is effective for defining CAN frame layouts and decoding rules and is supported by established industrial tools. For CAN-oriented projects, this makes DBC a practical and successful solution. At the same time, it remains tied to the CAN representation paradigm and does not provide a unified description basis beyond that domain.

\subsubsection{FIBEX}

The FIBEX \cite{s13} standard extends this idea toward broader automotive network description. ASAM defines FIBEX as a uniform XML-based interface description for configuring automotive network software, with extensions for CAN, LIN, FlexRay, and Ethernet. This makes FIBEX broader than DBC and highly relevant for multi-network automotive environments. However, it is still oriented toward a particular engineering ecosystem, a limited set of underlying protocols, and an associated set of implicit assumptions, rather than toward a general protocol-agnostic model for arbitrary heterogeneous serial communication formats.

\subsubsection{LIN Configuration Language (LDF)}

LDF \cite{s14} is another example of a useful but protocol-specific description approach. It provides a formal way to describe LIN networks and their configuration. This is valuable within the LIN domain, but it does not generalize into a universal basis for heterogeneous message description across unrelated protocol families.

\subsubsection{Summary}

Taken together, these formats and frameworks show that powerful automation-oriented solutions already exist and are successful within their intended domains. The purpose of the present work is not to replace them, but to provide a unified message model that can serve as a common basis across heterogeneous communication environments, including cases that fall outside the scope of domain-specific toolchain standards and protocol-specific description formats.

\subsection{Related Work}

The reviewed literature consistently indicates that modern embedded and avionics systems are becoming increasingly complex and increasingly difficult to develop efficiently. This complexity is associated not only with the growing number of integrated functions and devices, but also with the increasing number of engineering artifacts, configuration parameters, interfaces, and development dependencies that must remain mutually consistent throughout the process \cite{007,032,052}. In such projects, development effort is determined not only by the size of the system, but also by the diversity of artifacts and by the need to preserve consistency across multiple stakeholders, tools, and development stages.

To cope with this complexity, many authors propose model-based systems engineering (MBSE) and related digital-development approaches. In this line of work, models are treated as the central formal basis for organizing development information, improving traceability, and reducing inconsistencies between engineering artifacts \cite{029,038,048,051}. Such approaches are used to support early design activities, architectural refinement, validation planning, and communication between different engineering domains. In this sense, MBSE provides an important general framework for managing complexity, but by itself it does not yet guarantee efficient development.

For this reason, a number of authors go further and propose automation-oriented methods and toolchains. Halle and Thielecke describe AvioNET as a highly automated digital process framework for avionics platform development, emphasizing data continuity and interconnection of engineering activities \cite{038}. Brahmi et al.\ propose a formalized and automation-oriented software-engineering process for avionics, with the goal of improving safety and efficiency of development \cite{037,040}. Mueller et al.\ introduce a dedicated platform level for IMA systems development with extensive automation tool support \cite{048}. Panchal et al.\ describe a process-oriented build tool for airborne safety-critical software development \cite{049}. Hochstrasser proposes a modular model-based development approach for safety-critical flight-control software \cite{051}. Rhein et al.\ introduce a holistic interface-database approach in which communication data are stored in a central machine-readable form and transformed for downstream engineering tasks such as integration and test-bench development \cite{041}. Chrysalidis and Thielecke propose UCoF as a single-source-of-truth configuration format for virtual, hybrid, and hardware-based testing environments, with emphasis on configuration reuse and multi-fidelity development rather than on detailed formalization of message structure itself \cite{044}. Annighoefer et al.\ also describe automation-oriented platform modeling approaches in which ICDs and network-message-related artifacts are generated from a shared formal engineering basis \cite{026}. Taken together, these works show that MBSE becomes practically effective only when complemented by explicit automation mechanisms, shared machine-readable artifacts, and tool support for transformation and consistency maintenance.

At the same time, the reviewed work also shows that the question of how message data themselves should be represented and processed in a machine-readable form remains insufficiently addressed. Most automation-oriented approaches focus on architecture, platform configuration, validation workflows, or interface databases as broader engineering constructs. Comparatively close exceptions are the works of Watkins et al., which include explicit data-message modeling in the eSAM approach, but the presented solution is tied to ARINC~429-oriented representation and does not directly generalize across heterogeneous protocol families without introducing additional protocol-specific modeling branches \cite{001,003}. Thus, although the literature clearly demonstrates the value of MBSE and development automation, it still lacks a unified message model intended specifically for explicit and machine-processable description of heterogeneous transport-layer message structures. Such a model would not only address the problem considered in the present paper, but would also strengthen the automation-oriented approaches proposed in the works reviewed above.

\section{Message Model Proposal}

\subsection{Scope and Problem Statement}

The objective of the proposed model is to provide a unified representation of message formats used in heterogeneous serial communication environments. The focus of this work is limited to the structural description of transmitted messages.

Low-level transmission mechanisms---such as bit stuffing in CAN or byte framing conventions in UART interfaces (e.g., 8-N-1)---are outside the scope of the proposed model. Likewise, the model does not address higher-level communication procedures such as handshake protocols, session management, or transport-level exchanges (e.g., TCP connection handling). These aspects are assumed to be handled by lower-level transmission mechanisms or higher-level communication protocols surrounding the message structure.

Instead, the proposed approach focuses exclusively on the structural definition of message contents and their encoding.

The model is intended to support a wide range of serial communication technologies commonly encountered in embedded and avionics systems. Representative examples include, but are not limited to:

\begin{itemize}
\item ARINC 429
\item CAN BFF
\item CAN EFF
\item ARINC 825
\item UART-based protocols like MAVLink
\item UDP/IP/Ethernet stack
\end{itemize}
To ensure practical applicability across heterogeneous communication environments, the proposed message model approach is guided by a set of the following requirements:

\noindent\textbf{R1 --- Protocol universality}

The model must be capable of representing message formats used across heterogeneous communication protocols, including both standardized and custom serial message structures.

\noindent\textbf{R2 --- Complete message specification}

A complete message instance must be constructible solely from information contained in the model. The model must contain sufficient information to deterministically construct the exact binary representation of a message. No implicit structural assumptions or externally defined message elements should be required. This includes explicit representation of encoding semantics such as byte ordering, bit ordering, element positioning, and container sizes.

\noindent\textbf{R3 --- Data representation flexibility}

The model must not impose artificial or standard-driven restrictions on data types, addressing schemes, or value representations used within message structures.

The following subsections present the model developed to satisfy the stated requirements.
\subsection{Message Model}

\subsubsection{General data}

\begin{figure}[!htbp]
\centering
\includegraphics[width=\textwidth]{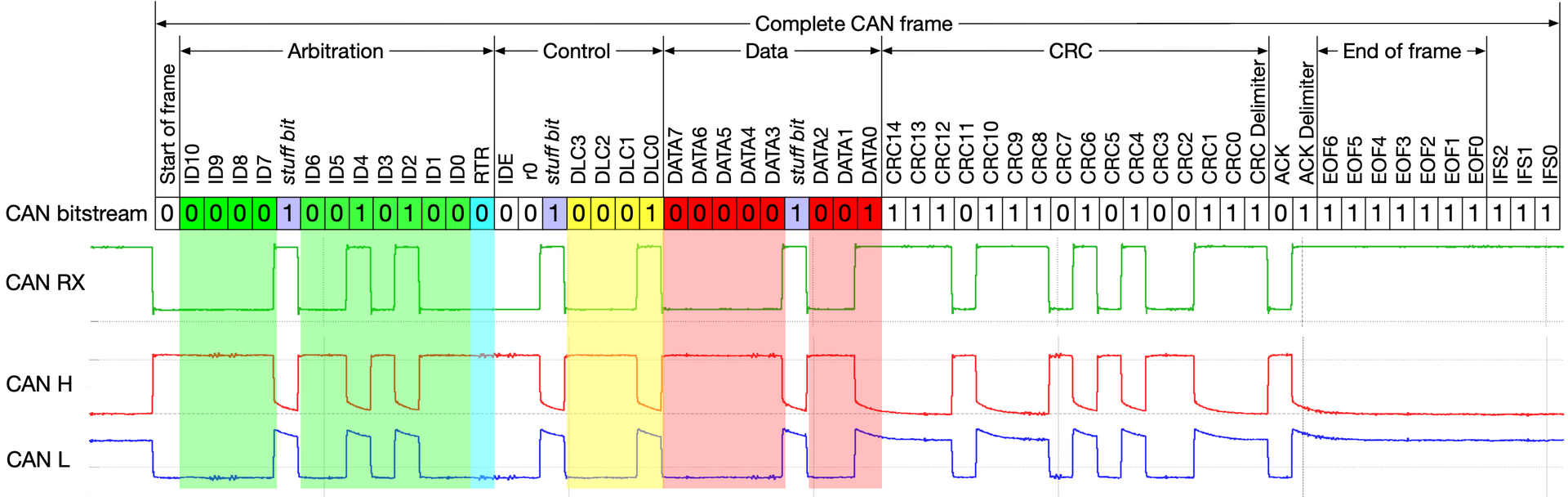}
\caption{CAN frame structure raw}
\label{fig:421_CAN_Message_raw}
\end{figure}

A typical message targeted by the proposed model can be illustrated by a representative raw CAN frame waveform, as shown in Fig.~\ref{fig:421_CAN_Message_raw}.

\begin{figure}[!htbp]
\centering
\includegraphics[width=\textwidth]{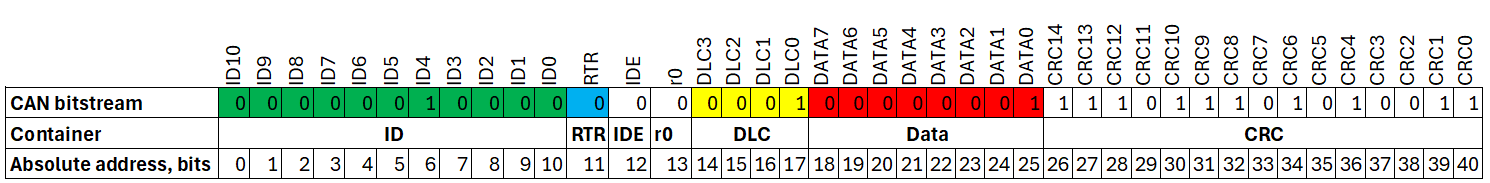}
\caption{CAN frame structure}
\label{fig:421_CAN_Message_structure}
\end{figure}

After removal of low-level transmission artifacts, such as start- and end-of-frame markers, inter-frame spacing, and stuff bits, the underlying message structure can be identified as shown in Fig.~\ref{fig:421_CAN_Message_structure}.

We call an atomic element of a message a \textit{container}. Each container corresponds to a specific fragment of the transmitted bitstream and is associated with a particular meaning within the protocol structure, such as identifier fields, payload data, control information, or checksum. It is also important to note that, in practice, message containers may represent data in a wide variety of formats. These formats differ in size, encoding rules, numeric interpretation, and structural composition. Therefore, before introducing the full message model, it is necessary to define the categories of data types that may appear within message containers.

Accordingly, the complete message model is structured into three parts, organized in a bottom-up manner:

\begin{itemize}
\item Data Type Model
\item Container Model
\item Message Model
\end{itemize}
\subsubsection{Data Type Model}

In the proposed approach, the set of available data types is not fixed. Users are allowed to define and extend the set of data types according to the needs of a specific project. However, all user-defined data types must belong to one of the fundamental data type groups described below:

\begin{itemize}
\item Boolean values
\item Integer values
\item Floating-point values
\item Fixed-point values
\item Binary-coded decimals (BCD)
\item Structured types
\end{itemize}
The majority of other commonly used representations can be expressed using these groups. For example, characters can be represented as unsigned 8-bit integers, strings as structures or arrays of characters, and opaque message fragments as containers with unspecified internal structure.

The model can also be easily extended to additional data type groups, for example Gray code (reflected binary code).

Each data type is characterized by a set of properties describing how the corresponding value is represented in the message. These properties define both the structural layout and the interpretation of encoded values.

Table~\ref{tab:dt_properties} summarizes the properties associated with data type.

\begin{table}[htbp]
\caption{Data type properties}\label{tab:dt_properties}
\begin{tabular}{p{0.16\textwidth}p{0.5\textwidth}p{0.2\textwidth}}
\toprule
Property & Description & Applicable groups \\
\midrule
Name & Unique identifier of the data type & all \\
Type group & One of the fundamental groups listed above & all \\
Bit length & Number of bits occupied by the value. May be unspecified at the type level. & all \\
Signedness & Indicates whether the value is signed or unsigned. May be unspecified at the type level. & integer, fixed-point \\
Sign encoding & Representation method used for signed values of integer data types (two's complement, one's complement, sign bit) & integer, fixed-point \\
Resolution & Value represented by the least significant bit & fixed-point, BCD \\
Offset & Offset applied when converting encoded values to physical quantities & fixed-point, BCD \\
\botrule
\end{tabular}
\end{table}

Not all properties are applicable to every data type group. For example, only fixed-point values require additional scaling parameters such as resolution and offset.

To support flexible type reuse, some properties of a data type may remain unspecified at the data type definition level. Such properties must be explicitly defined when the type is used within a message container. For example, an integer data type may have an unspecified size. In that case, the corresponding container must not only reference the data type, but also explicitly define the size property. This ensures that the exact binary layout of the message can be deterministically derived from the model.

\subsubsection{Container Model}

As mentioned above, a message is composed of structural elements referred to as containers. Each container represents a fragment of the transmitted message and defines how a particular piece of information is encoded within the message structure.

\paragraph{Container Roles}

The proposed model distinguishes several container roles according to their purpose within a message:

\noindent\textbf{Payload containers}

Payload containers carry application-level data. These containers encode engineering parameters exchanged between communicating systems. In many cases, payload containers represent the primary functional content of the message. Depending on the protocol, additional identifiers or control information may also appear within the payload.

\noindent\textbf{Identifier containers}

Identifier containers are used to distinguish message types or message instances within a communication channel. Such containers typically carry fixed values that allow receivers to recognize the message format and interpret their contents correctly.

\noindent\textbf{Length containers}

Length containers represent the total size of the transmitted message or payload fragment. These containers are required in protocols where message length is explicitly encoded as part of the transmitted frame (e.g. CAN).

\noindent\textbf{Checksum containers}

Checksum containers provide message integrity verification. Their values are calculated from other parts of the message according to a specified algorithm.

\noindent\textbf{Other containers}

Other containers represent message elements that do not fall into the categories above. In practice, such fields may include reserved areas, protocol-specific control bits, or placeholders intended for future use. Although they may not carry meaningful application data, they can still be represented to ensure accurate reconstruction of the transmitted message.

\paragraph{Container Properties}

Each container is defined by a set of properties (Tab.~\ref{tab:3} describing its role, structure, and encoding within the message.

\begin{table}[htbp]
\caption{Properties of a container}\label{tab:3}
\begin{tabular}{p{0.22\textwidth}p{0.72\textwidth}}
\toprule
Property & Description \\
\midrule
Name & Identifier of the container within the message definition \\
Role & Functional role of the container (payload, identifier, length, checksum, other) \\
Address & Absolute bit offset of the container relative to the beginning of the message \\
Data type & Reference to the data type used to interpret container contents \\
Byte endianness & Byte transmission order. Can be either "Big Endian" (Motorola) or "Little Endian" (Intel) \\
Bit endianness & Bit transmission order. Can be either "Big Endian" (most significant bit first) or "Little Endian" (least significant bit first). It is important to note that even though bit endianness normally match byte endianness it isn't always true. \\
Constant value & Fixed value associated with identifier containers \\
Implementation attributes & Parameters describing protocol-specific behavior \\
\botrule
\end{tabular}
\end{table}

The key requirement is that the binary representation of every container must be fully and explicitly defined by the model.

\subsubsection{Message Model}

A message is defined as a structured composition of containers arranged within a common bit-addressed space. The message model combines a set of global properties with a collection of containers that together define the complete binary representation of the transmitted data.

\paragraph{Basic Message Properties}

Each message is characterized by the properties specified in Table~\ref{tab:4}:

\begin{table}[htbp]
\caption{Properties of a message}\label{tab:4}
\begin{tabular}{p{0.22\textwidth}p{0.72\textwidth}}
\toprule
Property & Description \\
\midrule
Name & Unique human-readable identifier of the message \\
Length & Total length of the message in bits \\
Direction & Indicates whether the message is transmitted or received \\
Refresh rate & Transmission frequency for periodically transmitted messages \\
\botrule
\end{tabular}
\end{table}

\noindent\textit{Note: In practical applications, messages are typically defined within the context of predefined message types. A message type specifies allowable structures, constraints, and required elements for a class of messages. This includes restrictions on message size, container roles, and encoding rules. However, this concept doesn't affect the message model itself. Thus, it is introduced in detail in Section 5 as part of methods to work with the message model.}

\paragraph{User-Defined Message Properties}
In addition to the basic message properties defined for every message, the model allows users to introduce additional properties capturing message-type-specific behavior, such as the CRC calculation method or trailing-zero truncation as used in MAVLink. To keep the description focused, this paper does not further elaborate on this mechanism.

\paragraph{Message Structure Definition}

The internal structure of a message is defined as a collection of containers placed within the message bit space. Each container occupies a specific region defined by its absolute bit offset and size. The complete structure of the message is therefore fully determined by the set of containers and their associated properties. The following constraints apply

\begin{itemize}
\item containers must not overlap;
\item all containers must be fully contained within the message boundaries;
\item the combination of all containers must provide a complete and unambiguous description of the message structure.
\end{itemize}
\subsection{Implementation Examples}

This section presents representative examples demonstrating how the proposed message model can be applied to specific communication protocols.

The purpose of these examples is to illustrate how message structures can be described using the unified modeling approach. In each case, a message is represented as a collection of containers defined over a common bit-addressed space. Each container is assigned an explicit role, data type, and absolute bit offset within the message.

\clearpage

\subsubsection{ARINC 429}

An ARINC 429 message consists of a fixed-length 32-bit word containing multiple logically distinct fields. Using the proposed model, such a message can be represented as a set of containers defined over a shared bit-addressed structure as shown in Table~\ref{tab:example_A429}.

It should be noted that, according to the ARINC 429 standard, signed parameters are traditionally encoded by placing the sign bit in the most significant bit of the data field (in this example, bit 28), rather than using the more common two's complement representation. A strictly conventional ARINC 429 interpretation would therefore treat this bit as part of the SSM field. In the proposed model, however, this bit is intentionally retained as part of the payload container, resulting in a simpler and more consistent representation of the encoded parameter.

This modeling choice does not apply universally. For example, in cases involving SSM matrices of BCD type, separation of the sign from the matrix structure must be handled explicitly. The proposed model supports such cases, but requires a different container decomposition.

\begin{table}[!htbp]
\caption{ARINC 429 message example}
\label{tab:example_A429}
\centering
\scriptsize
\setlength{\tabcolsep}{2.5pt}
\renewcommand{\arraystretch}{1.08}

\begin{tabularx}{\textwidth}{|
>{\raggedright\arraybackslash}p{0.08\textwidth}|
>{\raggedright\arraybackslash}p{0.12\textwidth}|
>{\centering\arraybackslash}p{0.08\textwidth}|
>{\raggedright\arraybackslash}p{0.18\textwidth}|
>{\raggedright\arraybackslash}p{0.14\textwidth}|
>{\raggedright\arraybackslash}p{0.07\textwidth}|
X|}
\hline
\rowcolor{gray!20}
\textbf{Bits} & \textbf{Name} & \textbf{Role} & \textbf{Data Type} & \textbf{Endianness} & \textbf{Value} & \textbf{Comment} \\
\hline

0..7
& Label
& \cellcolor{idcell}ID
& int, unsigned, 8 bits
& byte: LE, bit: LSBF
& 212
& Decimal representation 212d stands for octal representation 324o. \\
\hline

8..9
& SDI
& \cellcolor{idcell}ID
& int, unsigned, 8 bits
& byte: LE, bit: LSBF
& 0
& Source Destination Identifier. \\
\hline

10..28
& A11\_Pitch
& \cellcolor{payloadcell}PL
& fixed-point, signed (sign bit, MSB), 19 bits, Res: 0.0006866455, Offset: 0
& byte: LE, bit: LSBF
& --
& Sign encoding is ``sign bit, MSB''. Specified resolution results in MSB weight of 90 degrees. \\
\hline

29..30
& SSM
& \cellcolor{payloadcell}PL
& int, unsigned, 2 bits
& byte: LE, bit: LSBF
& --
& Sign and Status matrix, defines the validity of transmitted data. \\
\hline

31
& ParityBit
& \cellcolor{checksumcell}CRC
& boolean
& N/A
& --
& -- \\
\hline

\end{tabularx}
\end{table}

\noindent\textit{Note: The examples in this section use the following acronyms and conventions:
ID -- identifier;
PL -- payload;
CRC -- cyclic redundancy check (and analogous integrity-check fields);
LEN -- length;
LSB/MSB -- least/most significant bit;
LE/BE -- little-/big-endian (byte order);
LSBF/MSBF -- least/most significant bit first;
Res. -- resolution.}

\clearpage

\subsubsection{CAN BFF}

Example in Table~\ref{tab:example_CAN_BFF} describes a CAN message with basic identifier and demonstrates several important aspects of the proposed modeling approach.

First, to ensure consistent interpretation of frame headers across platforms, all identifier-related containers must be explicitly defined, including RTR and IDE fields.

Second, although CAN is formally defined as a big-endian protocol, practical experience shows that commercial off-the-shelf (COTS) implementations may deviate from this convention. The proposed model allows such variations to be represented explicitly at the container level without introducing ambiguity into the message definition.

\begin{table}[!htbp]
\caption{CAN BFF message example}
\label{tab:example_CAN_BFF}
\centering
\scriptsize
\setlength{\tabcolsep}{2.5pt}
\renewcommand{\arraystretch}{1.08}

\begin{tabularx}{\textwidth}{|
>{\raggedright\arraybackslash}p{0.08\textwidth}|
>{\raggedright\arraybackslash}p{0.12\textwidth}|
>{\centering\arraybackslash}p{0.08\textwidth}|
>{\raggedright\arraybackslash}p{0.18\textwidth}|
>{\raggedright\arraybackslash}p{0.14\textwidth}|
>{\raggedright\arraybackslash}p{0.07\textwidth}|
X|}
\hline
\rowcolor{gray!20}
\textbf{Bits} & \textbf{Name} & \textbf{Role} & \textbf{Data Type} & \textbf{Endianness} & \textbf{Value} & \textbf{Comment} \\
\hline

0..10
& ID
& \cellcolor{idcell}ID
& int, unsigned, 11 bits
& byte: BE, bit: MSBF
& 12345
& Identifier. \\
\hline

11
& RTR
& \cellcolor{idcell}ID
& boolean
& N/A
& 0
& Must be dominant (0) for data frames and recessive (1) for remote request frames. \\
\hline

12
& IDE
& \cellcolor{idcell}ID
& boolean
& N/A
& 0
& Identifier extension bit (IDE). Must be dominant (0) for base frame format. \\
\hline

13
& r0
& \cellcolor{othercell}OTH
& boolean
& N/A
& 0
& Reserved bit. Must be dominant (0). \\
\hline

14..17
& DLC
& \cellcolor{lencell}LEN
& int, unsigned, 4 bits
& byte: N/A, bit: MSBF
& 6
& Data Length Code. \\
\hline

18..33
& A0Crnt
& \cellcolor{payloadcell}PL
& fixed-point, signed (2's compl.), 16 bits, Res.: 0.1, Offset: 0
& byte: LE, bit: MSBF
& --
& Even though CAN is big-endian by design, some data may be encoded as little-endian. \\
\hline

34..49
& A16ChEst
& \cellcolor{payloadcell}PL
& fixed-point, signed (2's compl.), 16 bits, Res.: 0.1, Offset: 0
& byte: LE, bit: MSBF
& --
& Even though CAN is big-endian by design, some data may be encoded as little-endian. \\
\hline

50..57
& A32TMin
& \cellcolor{payloadcell}PL
& fixed-point, unsigned, 8 bits, Res.: 1, Offset: -100
& byte: N/A, bit: MSBF
& --
& Resolution/offset combination results in a full range of -100..155 degrees. \\
\hline

58..65
& A40TMax
& \cellcolor{payloadcell}PL
& fixed-point, unsigned, 8 bits, Res.: 1, Offset: -100
& byte: N/A, bit: MSBF
& --
& Resolution/offset combination results in a full range of -100..155 degrees. \\
\hline

66..80
& CRC
& \cellcolor{checksumcell}CRC
& int, unsigned, 15 bits
& byte: BE, bit: MSBF
& --
& Checksum. \\
\hline

\end{tabularx}
\end{table}

\subsubsection{CAN EFF}

Example in Table~\ref{tab:example_CAN_EFF} presents a CAN message with extended identifier. A noteworthy aspect of this example concerns the representation of the extended CAN identifier. In engineering practice, it is common to treat the extended identifier as a single 29-bit value (in this example, 145358852 or 0x8AA0004). However, in the actual frame structure, the identifier is not transmitted as a single contiguous field. Instead, it is split into two separate components: IDA (11 bits) and IDB (18 bits), separated by control fields within the frame layout.

\begin{table}[!htbp]
\caption{CAN EFF message example}
\label{tab:example_CAN_EFF}
\centering
\scriptsize
\setlength{\tabcolsep}{2.5pt}
\renewcommand{\arraystretch}{1.08}

\begin{tabularx}{\textwidth}{|
>{\raggedright\arraybackslash}p{0.09\textwidth}|
>{\raggedright\arraybackslash}p{0.12\textwidth}|
>{\centering\arraybackslash}p{0.08\textwidth}|
>{\raggedright\arraybackslash}p{0.18\textwidth}|
>{\raggedright\arraybackslash}p{0.14\textwidth}|
>{\raggedright\arraybackslash}p{0.07\textwidth}|
X|}
\hline
\rowcolor{gray!20}
\textbf{Bits} & \textbf{Name} & \textbf{Role} & \textbf{Data Type} & \textbf{Endianness} & \textbf{Value} & \textbf{Comment} \\
\hline

0..10
& IDA
& \cellcolor{idcell}ID
& int, unsigned, 11 bits
& byte: BE, bit: MSBF
& 554
& First part of the identifier. Full identifier equals 145358852 (0x8AA0004) in the example. \\
\hline

11
& SRR
& \cellcolor{idcell}ID
& boolean
& N/A
& 1
& Substitute remote request. Must be recessive (1). \\
\hline

12
& IDE
& \cellcolor{idcell}ID
& boolean
& N/A
& 1
& Identifier extension bit (IDE). Must be recessive (1) for extended frame format with 29-bit identifiers. \\
\hline

13..30
& IDB
& \cellcolor{idcell}ID
& int, unsigned, 18 bits
& byte: BE, bit: MSBF
& 131076
& Second part of the identifier. \\
\hline

31
& RTR
& \cellcolor{idcell}ID
& boolean
& N/A
& 0
& Must be dominant (0) for data frames and recessive (1) for remote request frames. \\
\hline

32
& r1
& \cellcolor{othercell}OTH
& boolean
& N/A
& 0
& Reserved bit which must be set dominant (0), but accepted as either dominant or recessive. \\
\hline

33
& r0
& \cellcolor{othercell}OTH
& boolean
& N/A
& 0
& Reserved bit which must be set dominant (0), but accepted as either dominant or recessive. \\
\hline

34..37
& DLC
& \cellcolor{lencell}LEN
& int, unsigned, 4 bits
& byte: N/A, bit: MSBF
& 8
& Data Length Code. \\
\hline

38..69
& ActPos1
& \cellcolor{payloadcell}PL
& floating point single
& byte: BE, bit: MSBF
& --
& Payload element. \\
\hline

70..101
& ActPos2
& \cellcolor{payloadcell}PL
& floating point single
& byte: BE, bit: MSBF
& --
& Payload element.  \\
\hline

102..116
& CRC
& \cellcolor{checksumcell}CRC
& int, unsigned, 15 bits
& byte: BE, bit: MSBF
& --
& Checksum. \\
\hline

\end{tabularx}
\end{table}

The proposed model reflects the physical message structure rather than the conventional abstract representation. Accordingly, the identifier is described using two distinct containers (IDA and IDB), each defined at its corresponding bit offset. The complete identifier value is therefore reconstructed as: IDA $\times$ 2$^1$$^8$ + IDB.

This example highlights an important advantage of the proposed modeling approach: it preserves the exact structural organization of transmitted data while still allowing higher-level interpretations to be derived where needed.

\subsubsection{ARINC 825}
ARINC 825 adopts the CAN extended frame format but redefines the interpretation of the header and introduces additional constraints on allowed data types and their placement. For this reason, a separate structural example is not provided, as the CAN EFF example in Table~\ref{tab:example_CAN_EFF} already reflects the underlying message structure.

However, in practical applications, users may need to work with ARINC 825-specific abstractions and notation. The proposed modeling approach supports such use cases through configurable user representations, which allow alternative interpretations of the same underlying message structure. This capability is described in Sections 5 and 6.

\clearpage

\subsubsection{UART Serial Message Example \#1}

This example, taken from practical applications, presents a message with a simple header consisting of an identifier and a cyclic message counter, followed by a payload composed of 12-bit signed values and a trailing CRC field.

The example demonstrates that the proposed model is capable of representing non-standard and implementation-specific message formats with arbitrary field sizes and encoding rules.

\begin{table}[!htbp]
\caption{UART serial message example \#1}
\label{tab:example_UART1}
\centering
\scriptsize
\setlength{\tabcolsep}{2.5pt}
\renewcommand{\arraystretch}{1.08}

\begin{tabularx}{\textwidth}{|
>{\raggedright\arraybackslash}p{0.1\textwidth}|
>{\raggedright\arraybackslash}p{0.14\textwidth}|
>{\centering\arraybackslash}p{0.08\textwidth}|
>{\raggedright\arraybackslash}p{0.18\textwidth}|
>{\raggedright\arraybackslash}p{0.14\textwidth}|
>{\raggedright\arraybackslash}p{0.07\textwidth}|
X|}
\hline
\rowcolor{gray!20}
\textbf{Bits} & \textbf{Name} & \textbf{Role} & \textbf{Data Type} & \textbf{Endianness} & \textbf{Value} & \textbf{Comment} \\
\hline

0..31
& Header
& \cellcolor{idcell}ID
& int, unsigned, 32 bits
& byte: BE, bit: MSBF
& 16,777
& Message Identifier. \\
\hline

32..39
& Step
& \cellcolor{othercell}OTH
& int, unsigned, 8 bits
& byte: N/A, bit: MSBF
& --
& Message counter. \\
\hline

40..51
& D01\_LSLR
& \cellcolor{payloadcell}PL
& int, signed (2's compl.), 12 bits
& byte: BE, bit: MSBF
& --
& Payload element. \\
\hline

52..63
& D02\_LSFB
& \cellcolor{payloadcell}PL
& int, signed (2's compl.), 12 bits
& byte: BE, bit: MSBF
& --
& Payload element. \\
\hline

\multicolumn{7}{|c|}{\dots} \\
\hline

228..239
& D20\_POTR
& \cellcolor{payloadcell}PL
& int, signed (2's compl.), 12 bits
& byte: BE, bit: MSBF
& --
& Payload element. \\
\hline

240..255
& CRC
& \cellcolor{checksumcell}CRC
& int, unsigned, 16 bits
& byte: BE, bit: MSBF
& --
& Checksum. \\
\hline

\end{tabularx}
\end{table}
\clearpage

\subsubsection{UART Serial Message Example \#2 MAVLink}

Table~\ref{tab:example_MAVLink}  presents a MAVLink 2 standard message ATTITUDE (30) with no signature. The example is taken from the official web documentation \cite{s15} 

\begin{table}[!htbp]
\caption{MAVLink message example}
\label{tab:example_MAVLink}
\centering
\scriptsize
\setlength{\tabcolsep}{2.5pt}
\renewcommand{\arraystretch}{1.08}

\begin{tabularx}{\textwidth}{|
>{\raggedright\arraybackslash}p{0.09\textwidth}|
>{\raggedright\arraybackslash}p{0.17\textwidth}|
>{\centering\arraybackslash}p{0.06\textwidth}|
>{\raggedright\arraybackslash}p{0.18\textwidth}|
>{\raggedright\arraybackslash}p{0.12\textwidth}|
>{\centering\arraybackslash}p{0.07\textwidth}|
X|}
\hline
\rowcolor{gray!20}
\textbf{Bits} & \textbf{Name} & \textbf{Role} & \textbf{Data Type} & \textbf{Endianness} & \textbf{Value} & \textbf{Comment} \\
\hline

0..7
& STX
& \cellcolor{idcell}ID
& int, unsigned, 8 bits
& byte: N/A, bit: LSBF
& 253
& Protocol-specific start-of-text (STX) marker. Value 253 equals 0xFD. \\
\hline

8..15
& LEN
& \cellcolor{lencell}LEN
& int, unsigned, 8 bits
& byte: N/A, bit: LSBF
& 28
& Payload length. \\
\hline

16..23
& Incompatibility Flags
& \cellcolor{idcell}ID
& int, unsigned, 8 bits
& byte: N/A, bit: LSBF
& 0
& Flags that must be understood for MAVLink compatibility; implementation discards the packet if it does not understand the flag. \\
\hline

24..31
& Compatibility Flags
& \cellcolor{idcell}ID
& int, unsigned, 8 bits
& byte: N/A, bit: LSBF
& 0
& Flags that can be ignored if not understood; implementation may still process the packet. \\
\hline

32..39
& Packet sequence number
& \cellcolor{othercell}OTH
& int, unsigned, 8 bits
& byte: N/A, bit: LSBF
& 198
& Used to detect packet loss. Components increment this value for each transmitted message. \\
\hline

40..47
& System ID
& \cellcolor{idcell}ID
& int, unsigned, 8 bits
& byte: N/A, bit: LSBF
& 1
& ID of the system (vehicle) sending the message. \\
\hline

48..55
& Component ID
& \cellcolor{idcell}ID
& int, unsigned, 8 bits
& byte: N/A, bit: LSBF
& 1
& ID of the component sending the message. \\
\hline

56..79
& Message ID
& \cellcolor{idcell}ID
& int, unsigned, 24 bits
& byte: LE, bit: LSBF
& 30
& ID of the message type in payload. \\
\hline

80..111
& time\_boot\_ms
& \cellcolor{payloadcell}PL
& int, unsigned, 32 bits
& byte: LE, bit: LSBF
& --
& Timestamp (time since system boot) in milliseconds. \\
\hline

112..143
& roll
& \cellcolor{payloadcell}PL
& floating point single
& byte: LE, bit: LSBF
& --
& Roll angle in radians. \\
\hline

\multicolumn{7}{|c|}{\dots} \\
\hline

272..303
& yawspeed
& \cellcolor{payloadcell}PL
& floating point single
& byte: LE, bit: LSBF
& --
& Yaw angular speed in radians per second. \\
\hline

304..319
& Checksum
& \cellcolor{checksumcell}CRC
& int, unsigned, 16 bits
& byte: LE, bit: LSBF
& --
& CRC-16/MCRF4XX for the message. \\
\hline

\end{tabularx}
\end{table}

\subsubsection{UDP}

Table~\ref{tab:example_UDP}  illustrates a message whose header corresponds to the Ethernet--IP--UDP protocol stack. As shown in the table, the header occupies 42 bytes and consists of multiple fields defined across different protocol layers.

At first glance, such an explicit definition may appear excessive. However, this level of detail is required to satisfy the requirement R2 --- Complete message specification, ensuring that the message structure is fully and unambiguously defined.

The presence of a complete structural description does not imply that all elements must be actively used by all consumers of the model. For example, certain header fields may not be explicitly set in generated code. Nevertheless, their inclusion guarantees that all required information is available for any component that depends on it.

A practical challenge associated with such complete definitions is usability. This issue is addressed through configurable user representations, which allow the model to be presented and edited in a more convenient form. Details are provided in Sections 5 and 6.

\begin{table}[!htbp]
\caption{UDP message example}
\label{tab:example_UDP}
\centering
\scriptsize
\setlength{\tabcolsep}{2.5pt}
\renewcommand{\arraystretch}{1.08}

\begin{tabularx}{\textwidth}{|
>{\raggedright\arraybackslash}p{0.09\textwidth}|
>{\raggedright\arraybackslash}p{0.18\textwidth}|
>{\centering\arraybackslash}p{0.06\textwidth}|
>{\raggedright\arraybackslash}p{0.12\textwidth}|
>{\raggedright\arraybackslash}p{0.13\textwidth}|
>{\raggedright\arraybackslash}p{0.07\textwidth}|
X|}
\hline
\rowcolor{gray!20}
\textbf{Bits} & \textbf{Name} & \textbf{Role} & \textbf{Data Type} & \textbf{Endianness} & \textbf{Value} & \textbf{Comment} \\
\hline

0..47
& MAC\_Destination
& \cellcolor{idcell}ID
& int, uns., 48 bits
& byte: BE, bit: LSBF
& 0
& Destination MAC address. \\
\hline

48..95
& MAC\_Source
& \cellcolor{idcell}ID
& int, uns., 48 bits
& byte: BE, bit: LSBF
& 0
& Source MAC address. \\
\hline

96..111
& Ethertype
& \cellcolor{idcell}ID
& int, uns., 16 bits
& byte: BE, bit: LSBF
& 2048
& Value 0x0800 stands for IPv4. \\
\hline

112..115
& IP\_Version
& \cellcolor{idcell}ID
& int, uns.,  4 bits
& byte: N/A, bit: LSBF
& 4
& Always 4 for IPv4. \\
\hline

116..119
& IP\_IHL
& \cellcolor{idcell}ID
& int, uns.,  4 bits
& byte: N/A, bit: LSBF
& 5
& Internet Header Length. \\
\hline

120..125
& IP\_DSCP
& \cellcolor{idcell}ID
& int, uns., 6 bits
& byte: N/A, bit: LSBF
& 0
& Differentiated Services Code Point. \\
\hline

126..127
& IP\_ECN
& \cellcolor{othercell}OTH
& int, uns.,  2 bits
& byte: N/A, bit: LSBF
& 0
& Explicit Congestion Notification. \\
\hline

128..143
& IP\_TotalLen
& \cellcolor{lencell}LEN
& int, uns., 16 bits
& byte: BE, bit: LSBF
& 156
& Total IP packet size in bytes \\
\hline

144..159
& IP\_Ident
& \cellcolor{othercell}OTH
& int, uns., 16 bits
& byte: BE, bit: LSBF
& 0
& Identification field for fragmented datagrams. \\
\hline

160..162
& IP\_Flags
& \cellcolor{othercell}OTH
& int, uns., 3 bits
& byte: N/A, bit: LSBF
& 0
& This feature is not considered.\\
\hline

163..175
& IP\_FrgmOffset
& \cellcolor{othercell}OTH
& int, uns., 13 bits
& byte: BE, bit: LSBF
& 0
& This feature is not considered. \\
\hline

176..183
& IP\_TTL
& \cellcolor{othercell}OTH
& int, uns., 8 bits
& byte: N/A, bit: LSBF
& 255
& This feature is not considered.  \\
\hline

184..191
& IP\_Protocol
& \cellcolor{idcell}ID
& int, uns., 8 bits
& byte: N/A, bit: LSBF
& 17
& Value 17 stands for UDP. \\
\hline

192..207
& IP\_HeaderCRC
& \cellcolor{checksumcell}CRC
& int, uns., 16 bits
& byte: BE, bit: LSBF
& --
& IPv4 header checksum \\
\hline

208..239
& IP\_SrcAddress
& \cellcolor{idcell}ID
& int, uns., 32 bits
& byte: BE, bit: LSBF
& 3,232, 235, 777
& IPv4 address of the sender. Value 3232235777 stands for ``192.168.1.1''. \\
\hline

240..271
& IP\_DstAddress
& \cellcolor{idcell}ID
& int, uns., 32 bits
& byte: BE, bit: LSBF
& 3, 232, 235, 778
& IPv4 address of the receiver. Value 3232235778 stands for ``192.168.1.2''. \\
\hline

272..287
& UDP\_SrcPort
& \cellcolor{idcell}ID
& int, uns., 16 bits
& byte: BE, bit: LSBF
& 49001
& Identifies the sender's UDP port. \\
\hline

288..303
& UDP\_DstPort
& \cellcolor{idcell}ID
& int, uns., 16 bits
& byte: BE, bit: LSBF
& 49002
& Identifies the receiver's UDP port. \\
\hline

304..319
& UDP\_Length
& \cellcolor{lencell}LEN
& int, uns., 16 bits
& byte: BE, bit: LSBF
& 136
& Length of the UDP datagram in bytes. \\
\hline

320..335
& UDP\_CRC
& \cellcolor{checksumcell}CRC
& int, uns., 16 bits
& byte: BE, bit: LSBF
& --
& UDP header and payload checksum. \\
\hline

336..399
& A0ActPosCh1
& \cellcolor{payloadcell}PL
& floating point double
& byte: BE, bit: LSBF
& --
& Payload element. \\
\hline

400..461
& A32ActPosCh2
& \cellcolor{payloadcell}PL
& floating point double
& byte: BE, bit: LSBF
& --
& Payload element. \\
\hline

\end{tabularx}
\end{table}

\subsection{Message Model Summary}

The proposed message model provides a fully explicit and deterministic description of message structure. All requirements formulated in Section 4.1 are satisfied: the model supports heterogeneous communication protocols, enables complete reconstruction of message binary representation, and does not impose artificial restrictions on data representation. The model achieves this by combining:

\begin{itemize}
\item a flexible data type system,
\item a container-based structural representation,
\item and a bit-accurate definition of message structure.
\end{itemize}
As a result, message definitions are unambiguous, implementation-independent, and suitable for automated processing.

At the same time, the explicit nature of the model introduces practical challenges. Direct manual creation and maintenance of message descriptions in this form may be labor-intensive and error-prone, especially for complex protocols or large interface definitions. To enable practical application of the proposed approach, additional methods and supporting tooling are required. In particular, it is necessary to provide:

\begin{itemize}
\item mechanisms for defining and enforcing structural constraints (e.g., message types);
\item automated creation of mandatory containers and structural elements;
\item user-friendly representations (user representations) for efficient inspection and editing of message definitions;
\item configurable user interfaces that allow adaptation of the model representation to domain-specific conventions (e.g., CAN vs. ARINC 825 interpretations).
\end{itemize}
These aspects are addressed in the following sections.

\section{Methods for Working with the Model}

\subsection{Problem Statement}

The message model introduced in Section 4 provides an explicit and fully defined representation of message structure. However, constructing and maintaining message descriptions directly at this level of detail is impractical in real engineering workflows. Manual specification of containers and their properties is time-consuming, error-prone, and difficult to scale, particularly when working with large interface datasets.

To enable effective practical use of the proposed model, structured methods are required for constructing, maintaining, and validating message descriptions. These methods must satisfy the following requirements.

\noindent\textbf{T1. Constraint-based consistency.}

The methods shall support maintaining consistency between message definitions and user-defined structural and semantic constraints.

\noindent\textit{Explanation. Such constraints may be derived from communication standards or from project-specific protocols, including cases where the corresponding rules are described only informally. The methods must therefore support both formalized and user-defined constraint sets.}

\noindent\textbf{T2. Support for deterministic automation.}

The methods shall support automatic creation and completion of all message elements and properties that can be derived unambiguously from existing model data, constraint definitions, or user-provided engineering-level properties.

\noindent\textit{Explanation. This includes, for example, automatic creation of mandatory containers and automatic assignment of their values when these values are uniquely determined by structural rules or by user-facing message properties.}

\noindent\textbf{T3. User-oriented model interaction.}

The methods shall support presentation and editing of the formal model in a form convenient and familiar to the user.

\noindent\textit{Explanation. The user should not be forced to work directly with the raw structural representation when a more natural engineering-oriented representation can be provided.}

\noindent\textbf{T4. User-configurable representations.}

The representation format used for model presentation and editing shall be configurable by the user and shall not be hard-coded.

\noindent\textit{Explanation. Different domains and workflows may require different engineering notations and abstractions. The methods must therefore allow the same formal model to be presented through different user-defined representations.}

\subsection{Implementation Basics}

The proposed implementation addresses the requirements stated above through two complementary mechanisms.

First, a message type mechanism is introduced. In the context of this work, a message type is a user-defined description of allowable message structure, including structural constraints, permitted container configurations, and mandatory elements that can be generated automatically. This mechanism provides the basis for consistency enforcement and support of deterministic automation.

Second, configurable user representations are introduced for messages and containers. These representations allow the formal model to be viewed and edited in a user-oriented form while preserving a bidirectional mapping to the underlying structural definition. This mechanism provides convenient engineering interaction with the model and enables adaptation of the interface to different domain-specific conventions.

Together, these two mechanisms address the full set of requirements formulated above: message types provide structured constraint handling and automation support, while configurable representations provide practical and user-adjustable access to the formal model. The message type mechanism is described in Section 5.3. The representation mechanism is described in Section 5.4.

\subsection{Message Types}

\subsubsection{Mechanism Description}

Message types define the allowable characteristics and structural constraints of messages within a given communication domain.

In practical workflows, message types serve as a primary mechanism for reducing manual effort and preventing structural inconsistencies. For example, when a user specifies that a message belongs to the CAN EFF type, this requirement uniquely determines the structure of the frame header and constrains the allowable payload organization. Message types therefore provide both structural guidance and validation capabilities for message definitions.

Message types define the following properties:

\begin{itemize}
\item Message properties constraints
\item Structural constraints
\end{itemize}
Message properties constraints specify allowable size limits expressed in bits.

Structural constraints define allowable message structures in terms of container types and their properties. For each container type, the following characteristics are specified:

\begin{enumerate}
\item Allowed address range. Range of permissible bit offsets for containers of this type
\item Container type name.
\item Container role. Functional role of the container described above.
\item Allowed data types. Either a defined set of permissible data types or an indication that any type may be used.
\item Allowed size range. Permissible bit-length range for containers of this type.
\item Minimum and maximum number of instances. The property specifying how many containers of this type may appear within a message.
\item Byte order (endianness). Defines the byte order used for the container. Can be either big- or little- endian.
\item Bit order. Defines the bit order. Can be either most-significant-bit-first or least-significant-bit-first.
\item Automatic creation flag. An indicator specifying whether containers of this type should be generated automatically when a message is created.
\end{enumerate}
Automatically generated containers are created during message instantiation based on the selected message type. Their primary purpose is to ensure consistent creation of mandatory structural elements such as identifier fields.

To enable automatic creation, all defining properties of a container must be unambiguous. In particular, the data type, size, and placement constraints must be fully determined by the message type definition. Only containers meeting these conditions can be generated automatically without introducing ambiguity into the message model.

\subsubsection{Example: CAN EFF Message Type Definition}

To illustrate the message type mechanism, this subsection presents a simplified CAN EFF message type definition derived from an actual implementation. The example shows how a message type captures both message-level constraints and structural rules governing container placement, size, encoding properties, and automatic creation.

For the CAN EFF message type, the allowable message size ranges from 53 to 117 bits depending on payload length. These limits reflect protocol-specific structural overhead and define the valid range for all compliant message instances.

The message structure is defined as a set of container types with explicitly constrained properties. Each container definition includes the properties specified in Section 5.3.1. The corresponding message type definition is shown in the table ~\ref{tab:example_CAN_EFF_structure}.

\begin{table}[!htbp]
\caption{CAN EFF message structure definition}
\label{tab:example_CAN_EFF_structure}
\centering
\scriptsize
\setlength{\tabcolsep}{2.5pt}
\renewcommand{\arraystretch}{1.08}

\begin{tabularx}{\textwidth}{|
>{\raggedright\arraybackslash}p{0.1\textwidth}|
>{\raggedright\arraybackslash}X|
>{\raggedright\arraybackslash}p{0.08\textwidth}|
>{\raggedright\arraybackslash}p{0.22\textwidth}|
>{\raggedright\arraybackslash}p{0.07\textwidth}|
>{\raggedright\arraybackslash}p{0.14\textwidth}|
>{\raggedright\arraybackslash}p{0.10\textwidth}|}
\hline
\rowcolor{gray!20}
\textbf{Address range} & \textbf{Name} & \textbf{Role} & \textbf{Data Type} & \textbf{Qty.} & \textbf{Endianness} & \textbf{Auto-Create} \\
\hline

0..10
& IDA
& \cellcolor{idcell}ID
& int, unsigned, 11 bits
& 1
& byte: BE, bit: MSBF
& yes \\
\hline

11
& SRR
& \cellcolor{idcell}ID
& boolean
& 1
& N/A
& yes \\
\hline

12
& IDE
& \cellcolor{idcell}ID
& boolean
& 1
& N/A
& yes \\
\hline

13..30
& IDB
& \cellcolor{idcell}ID
& int, unsigned, 18 bits
& 1
& byte: BE, bit: MSBF
& yes \\
\hline

31
& RTR
& \cellcolor{idcell}ID
& boolean
& 1
& N/A
& yes \\
\hline

32
& r1
& \cellcolor{othercell}OTH
& boolean
& 1
& N/A
& yes \\
\hline

33
& r0
& \cellcolor{othercell}OTH
& boolean
& 1
& N/A
& yes \\
\hline

34..37
& DLC
& \cellcolor{lencell}LEN
& int, unsigned, 4 bits
& 1
& byte: BE, bit: MSBF
& yes \\
\hline

38..101
& Payload\_BE
& \cellcolor{payloadcell}PL
& unrestricted, size: 1..64 bits
& 0..64
& byte: BE, bit: MSBF
& no \\
\hline

38..101
& Payload\_LE
& \cellcolor{payloadcell}PL
& unrestricted, size: 1..64 bits
& 0..64
& byte: LE, bit: MSBF
& no \\
\hline

trailing 0..14
& CRC
& \cellcolor{checksumcell}CRC
& int, unsigned, 15 bits
& 1
& byte: BE, bit: MSBF
& yes \\
\hline

\end{tabularx}
\end{table}

This example illustrates several important aspects of the proposed message type mechanism.

First, the complete message structure is defined explicitly through constrained container types rather than implicit protocol knowledge. This enables deterministic validation and consistency enforcement of message definitions and eliminates ambiguity in implementation.

Second, specification of data types for each container plays a fundamental role in the model. As defined in Section 4.2.2, data types determine encoding semantics and ensure consistent interpretation.

Third, encoding properties such as byte order and bit order are defined explicitly for each container type, ensuring that the structural model fully captures transmission semantics across implementations and architectures.

Fourth, mandatory structural elements such as identifiers, control fields, and CRC are configured for automatic creation, while flexible regions such as payload fields are defined through bounded ranges rather than fixed layouts.

In this example, the two payload container types ("Payload\_BE", "Payload\_LE") represent alternative allowable configurations (byte endianness) rather than simultaneously present elements of the same message instance. The CRC placement is defined relative to the end of the message because its absolute position depends on payload size.

\subsection{User Representation Method}

\subsubsection{Mechanism Description}

\paragraph{Core Idea}

The proposed approach introduces a two-level representation mechanism designed to make the detailed message model practical for everyday engineering use. The key idea is to separate the internal model representation from the user-facing representation while maintaining a well-defined functional relationship between them.

For both messages and containers, the representation is organized into two independently defined but functionally linked levels:

\begin{itemize}
\item Data-level representation, which corresponds directly to the underlying formal model;
\item User-level representation, which defines how the same information is presented and edited by engineers.
\end{itemize}
The data-level representation operates directly on model properties, while the user-level representation defines the structure and semantics of the interface visible to users.

The key property of this mechanism is its bidirectional nature: the same configuration defines both how model data is presented to the user and how user input is translated back into the formal model. The user representation is therefore not merely a visualization layer, but an operational interface for constructing and modifying the formal message description.

All mappings, visibility rules, and transformations should be fully configurable and defined independently of message type. This allows the same structural model to support different engineering conventions without introducing parallel message type definitions.

An important aspect of the proposed approach is that message-level representations may include not only properties of the message itself but also selected properties of nested containers, particularly those that are unique and automatically generated.

For example, identifier values are often exposed directly at the message level, while in the formal model they are represented as properties of containers with the "Identifier" role. Similarly, modifying a message-level property such as payload size may simultaneously update both the corresponding message attribute and nested container values.

Not all containers included in the full structural representation are relevant for routine user interaction. In practice, users often prefer to hide:

\begin{itemize}
\item identifier- or length-related containers as they are already exposed at the message level,
\item automatically generated elements such as CRC fields.
\end{itemize}
\paragraph{Core Idea Example: Payload-Oriented Size Representation}

The advantages of the proposed mechanism can be illustrated using a typical case where the formal model and engineering practice rely on different abstractions.

In the formal model, message size is defined as the total number of bits in the frame. In contrast, engineers often work with payload size expressed in bytes. Since the size of protocol overhead is fixed for a given message type, the user-level representation can define payload size as a derived property computed from the total message size. For example, in CAN EFF: SIZEgui = (SIZEmodel - 53)/8, where 53 is a size of overhead in bits.

The corresponding reverse transformation allows user-edited payload size values to be translated back into total frame size and associated structural elements. For example, in CAN EFF: SIZEmodel = SIZEgui*8+53

As a result, engineers can operate using familiar abstractions while the underlying model remains complete and structurally explicit.

\subsubsection{Use Case Example for CAN EFF}

To illustrate the proposed representation method, this section presents an example configuration for a CAN\_EFF message. The purpose of the example is not to describe the full configuration syntax, but rather to demonstrate the key principles of the method described above.

\paragraph{Message-Level Representation}

For the message type CAN\_EFF, the user-facing representation is configured independently of the underlying message model. At the message level, the user is presented with a compact set of engineering-oriented message properties, including properties derived from the underlying containers.

A simplified pseudoconfiguration is shown below:

\begin{verbatim}
Message representation: CAN_EFF

User-visible properties:
FID_dec  := ContainerValue(IDA)*2^18 + ContainerValue(IDB)***
FID_hex*  := DecToHex(ContainerValue(IDA)*2^18 + ContainerValue(IDB))***
DLC**      := (Size_bits - 53) / 8
Name*     := If(Description="", "F"+FID_dec, "F"+FID_dec+"_"+Description)
Direction := Direction
RefreshRate_ms := RefreshRate_ms
Description := Description
\end{verbatim}

\begin{verbatim}
*Read-only properties are marked with asterisk
**DLC represents the size of the payload in bytes, while Size_bits is 
the size of the complete message in bits. 53 bits is a sum of all 
identifiers, checksums etc.
*** ContainerValue is a function that returns the value of the nested 
container with the specified name.
\end{verbatim}

This representation demonstrates that the user-visible message view does not need to mirror the structural composition of the underlying frame. Instead, it can expose only those properties that are meaningful for routine engineering tasks. In particular:

\begin{itemize}
\item the extended CAN identifier is shown as a single value, although it is physically represented in the message as two separate containers, IDA and IDB;
\item the size shown to the user is the payload-oriented value DLC, rather than the total message size in bits;
\item naming can be dynamically derived from message properties rather than stored as a separate model element.
\end{itemize}
\paragraph{Message-Level Reverse Transformation}

The representation method is bidirectional. In addition to deriving user-facing properties from the model, it also defines how user input is translated back into raw model data. For the same example, the reverse transformations may be expressed as follows:

\begin{verbatim}
Message Reverse transformations: CAN_EFF

IDA := floor(FID_dec / 2^18)
IDB := FID_dec mod 2^18
Size_bits := 8*DLC + 53
DLC.value := DLC
Direction := Direction
RefreshRate_ms := RefreshRate_ms
Description := Description

\end{verbatim}

This illustrates two important cases. First, a single user-visible property may be mapped back into multiple model elements. For example, editing FID\_dec updates both IDA and IDB. Second, a user-visible property may simultaneously affect a message-level property and a nested container. For example, editing DLC updates the total message size as well as the value of the corresponding DLC container. This mechanism allows engineers to work with familiar abstractions while preserving a fully explicit and structurally accurate formal model underneath.

\paragraph{Container-Level Representation}

A separate representation is configured for payload containers. Here again, the user-facing view is adapted to engineering practice rather than to the raw bit-level message structure.

Since this representation is intended specifically for payload elements, non-payload structural containers such as identifier fields, length-related elements, and checksum fields are hidden from this view.

In the message model, container addresses are defined as absolute bit offsets from the beginning of the message. In the user-facing representation, however, addresses are displayed relative to the beginning of the payload area.

In this example, container names are also derived automatically: each name is composed of an address-based prefix and the container description.

A simplified pseudoconfiguration is shown below:

\begin{verbatim}
Container representation: CAN_EFF payload

Hidden containers:
IDA, IDB, CRC, SRR, RTR, IDE, r0, r1, DLC

Visible properties:
Address    := Address_model - 38
Name*       := If(Description="", "A"+Address, "A"+Address+"_"+Description)
DataType   := DataType
Size       := Size
Resolution := Resolution
Offset     := Offset
Sign       := Sign
Description := Description

*Read-only properties are marked with asterisk
\end{verbatim}

\paragraph{Container-Level Reverse Transformation}

The container-level reverse transformation is as simple as:

\begin{verbatim}
Address_model := Address + 38
DataType   := DataType
Size       := Size
Resolution := Resolution
Offset     := Offset
Sign       := Sign
\end{verbatim}

As a result, the engineer works with payload-relative coordinates, while the model retains the absolute addressing required for deterministic message construction, validation, and code generation.

\paragraph{Discussion}

This example demonstrates that the proposed representation method is not limited to passive formatting of model data. Instead, it provides a configurable, bidirectional transformation layer between the formal message model and the user-facing engineering representation. This makes it possible to:

\begin{itemize}
\item expose only relevant properties to the user,
\item hide automatically generated or structurally derived elements,
\item present data in forms consistent with protocol documentation and engineering conventions,
\item transform user input back into a complete and explicit message model.
\end{itemize}
An important implication of the proposed representation method is that different user-facing views can be configured for the same underlying frame structure.

A representative example is ARINC 825, which relies on the same underlying frame definition as CAN EFF. The primary differences arise from how the frame header is interpreted and represented: ARINC 825 conceptually decomposes the identifier field into separate components and imposes additional constraints on allowable payload data types and their placement. However, these differences do not affect the underlying structure of the frame.

Using the proposed representation method, CAN EFF and ARINC 825 can therefore share the same formal message model while employing different user representations. This demonstrates that the method separates the structural description of a message from its engineering interpretation and allows multiple domain-specific views to coexist over the same underlying data.

\section{Modelling Tool Example}

\subsection{Implementation Approach}

The proposed modeling approach has been implemented in an industrial MBSE tool environment dBricks \cite{t1}, which serves as a reference implementation of the concepts described in this paper.

In the implemented environment, configuration of message types and user representations follows directly the methodology introduced in Sections 4 and 5. Message type definitions specify structural constraints, container roles, encoding properties, and automatic-generation rules. Representation configurations define user-oriented views and bidirectional transformations between the formal model and engineering-facing abstractions.

As a result, the tool does not introduce additional modeling concepts beyond those described in this work; instead, it provides a practical environment for their consistent application in real development workflows.

\subsection{Message Creation and Editing Workflow}

\begin{figure}[!htbp]
\centering
\includegraphics[width=.8\textwidth]{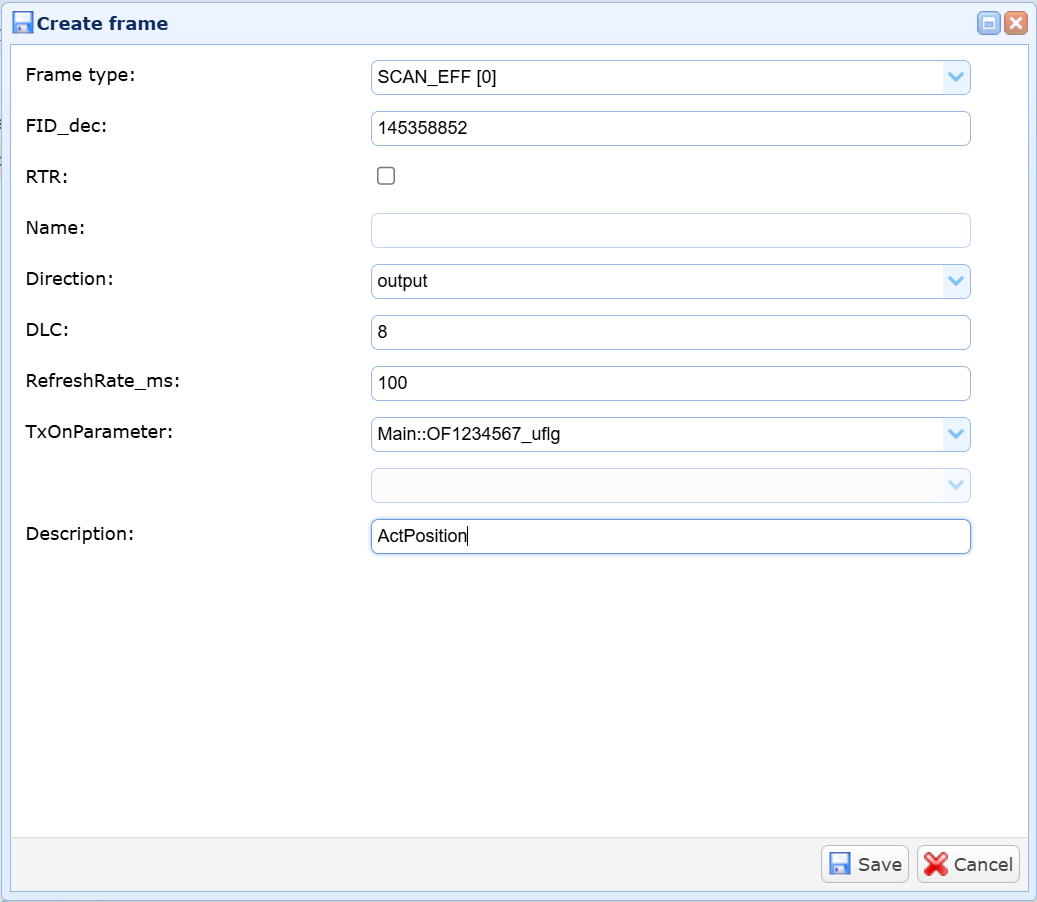}
\caption{"Create Frame" form for CAN EFF}
\label{fig:62a_MessageCreation}
\end{figure}

This example demonstrates how the proposed modeling approach operates in a real engineering workflow. Figure~\ref{fig:62a_MessageCreation} shows the message creation form used to define a new CAN extended frame (CAN EFF) message.

\noindent\textit{Note: dBricks SW uses the term "Frame" as synonym for "Message"}

At creation time, the user specifies only high-level engineering properties exposed by the configured representation. In this example, these include the message type, identifier value (FID\_dec), payload size (DLC), direction, refresh rate, transmission control parameter, and description.

Notably, no protocol-level structural elements are defined manually. The user does not specify container composition, addressing, encoding properties, or control fields such as IDA, IDB, or CRC. These elements are instead derived automatically from the selected message type and associated constraints.

Following message creation, the tool automatically generates a complete structural representation of the frame based solely on the selected message type.

\begin{figure}[!htbp]
\centering
\includegraphics[width=.8\textwidth]{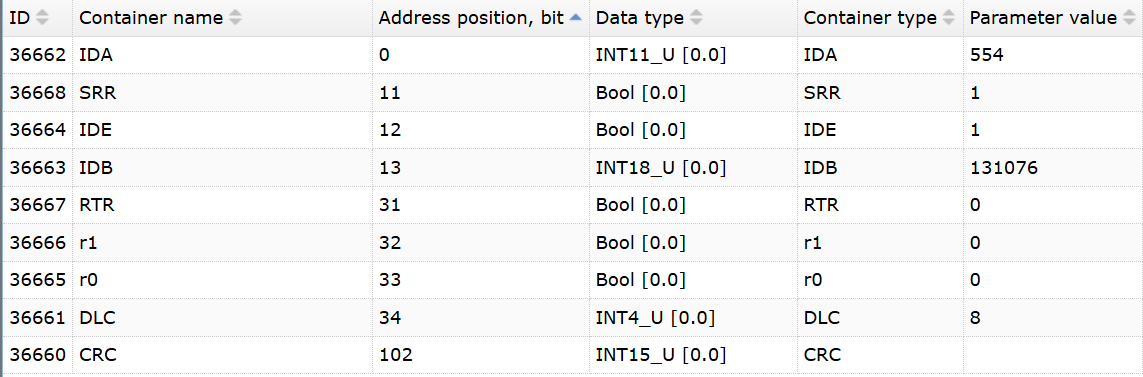}
\caption{CAN EFF raw structural representation after creation}
\label{fig:62b_MessageStructureRaw1}
\end{figure}

Figure~\ref{fig:62b_MessageStructureRaw1} shows the resulting structural definition immediately after message creation. All mandatory protocol-level containers---including IDA, IDB, reserved fields, DLC, and CRC---are generated automatically according to the configured constraints. The generated structure reflects the exact physical layout of the CAN EFF frame, including correct addressing of all elements and placement of the CRC container at the end of the message.

Values for identifier-related containers and the length container are assigned automatically. In this example, the identifier entered by the user is decomposed into IDA and IDB components, while the DLC container is assigned the corresponding payload size value.

In parallel with the structural representation, the tool provides a user-oriented view of the message definition derived from the same underlying model.

\begin{figure}[!htbp]
\centering
\includegraphics[width=.8\textwidth]{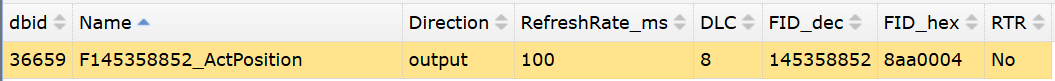}
\caption{CAN EFF message user representation after creation}
\label{fig:62c_MessageUser1}
\end{figure}

Figure~\ref{fig:62c_MessageUser1} shows the configured message-level representation of the same CAN EFF frame. Instead of exposing the full structural composition, this view presents a compact set of engineering-relevant properties, including the identifier value in both decimal and hexadecimal form, payload size (DLC), direction, and refresh rate.

Protocol-level structural elements such as IDA, IDB, control fields, and CRC remain hidden in this view, although they are fully defined in the underlying model. Despite the simplified presentation, all values shown here are directly derived from the formal message definition.

\begin{figure}[!htbp]
\centering
\includegraphics[width=.8\textwidth]{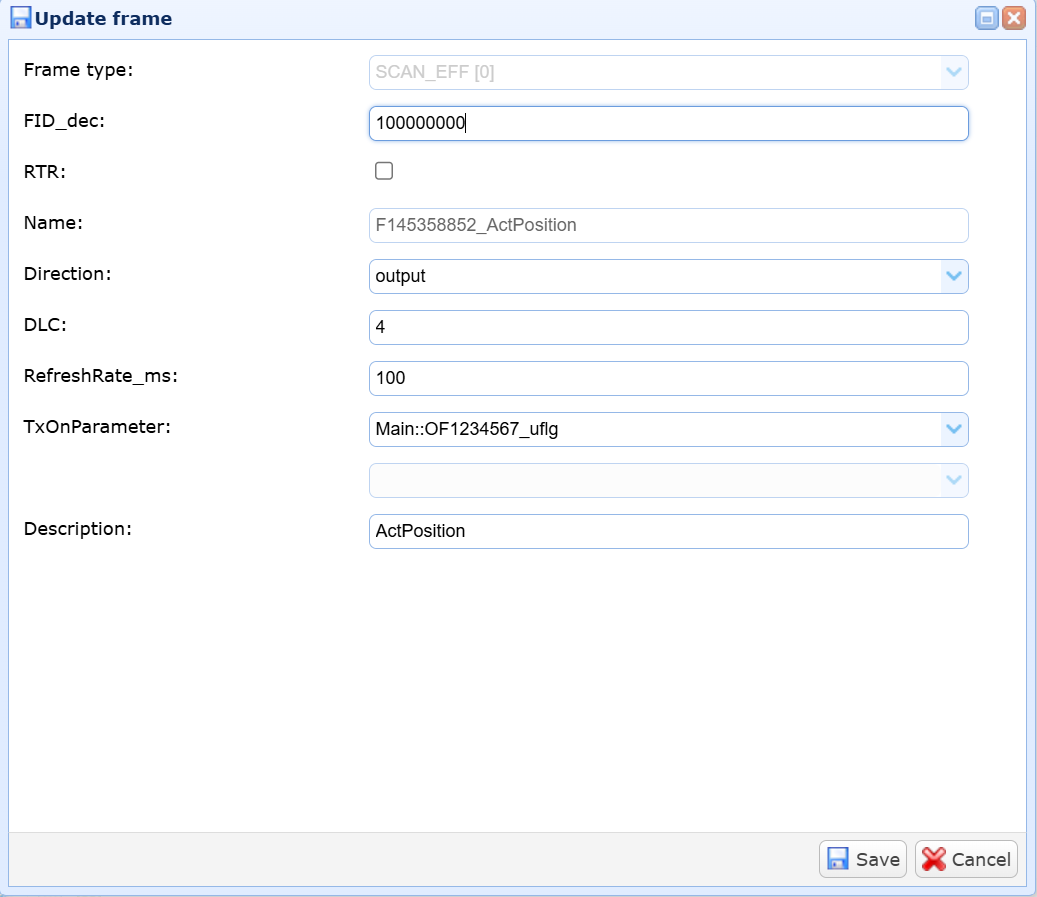}
\caption{"Update Frame" form for CAN EFF}
\label{fig:62d_MessageUpdateForm}
\end{figure}

Figure~\ref{fig:62d_MessageUpdateForm} shows the update form where both the identifier value and payload size (DLC) are modified by the user. As in the creation stage, only high-level engineering properties are edited, while protocol-level structural elements remain hidden from direct interaction.

These changes trigger automatic updates of the underlying structural representation.

\begin{figure}[!htbp]
\centering
\includegraphics[width=.8\textwidth]{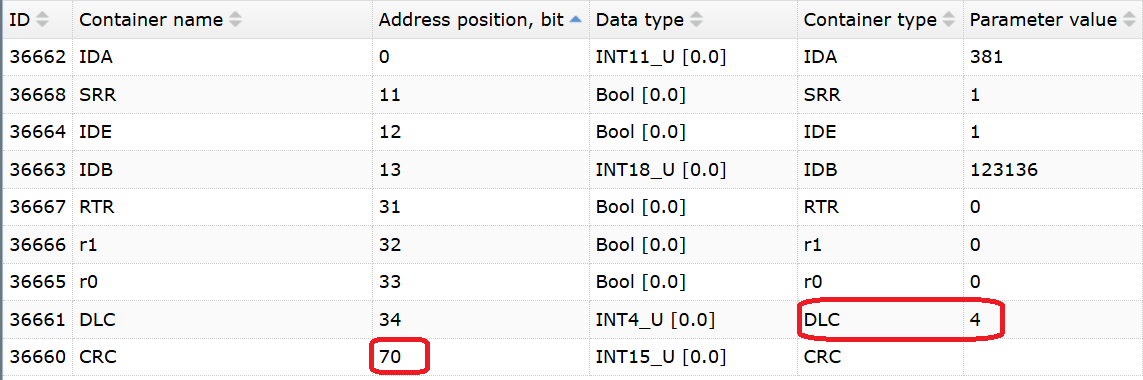}
\caption{CAN EFF raw structural representation after update}
\label{fig:62e_MessageStructureRaw2}
\end{figure}

Figure~\ref{fig:62e_MessageStructureRaw2} shows the updated structural representation after modifying both the identifier and payload size. The underlying model is automatically recalculated without any manual adjustments to protocol-level elements. The identifier value entered by the user is decomposed into updated IDA and IDB containers, while the DLC container reflects the new payload size. In addition, dependent structural elements are repositioned automatically: in particular, the CRC container is moved according to the new message length.

\begin{figure}[!htbp]
\centering
\includegraphics[width=.8\textwidth]{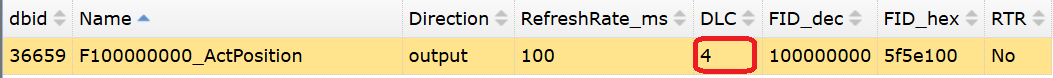}
\caption{CAN EFF message user representation after modification}
\label{fig:62f}
\end{figure}

The corresponding user-oriented representation is shown in Figure~\ref{fig:62f}. The updated identifier value is presented consistently in both decimal and hexadecimal form, and the payload size reflects the modified DLC value.

Together, these examples confirm that changes introduced through the user-facing interface are propagated deterministically to the raw message model, demonstrating the bidirectional consistency of the proposed approach.

\subsection{Payload Container Definition Workflow}

This subsection demonstrates how payload containers are defined and integrated into the message structure using the proposed modeling environment.

\begin{figure}[!htbp]
\centering
\includegraphics[width=.8\textwidth]{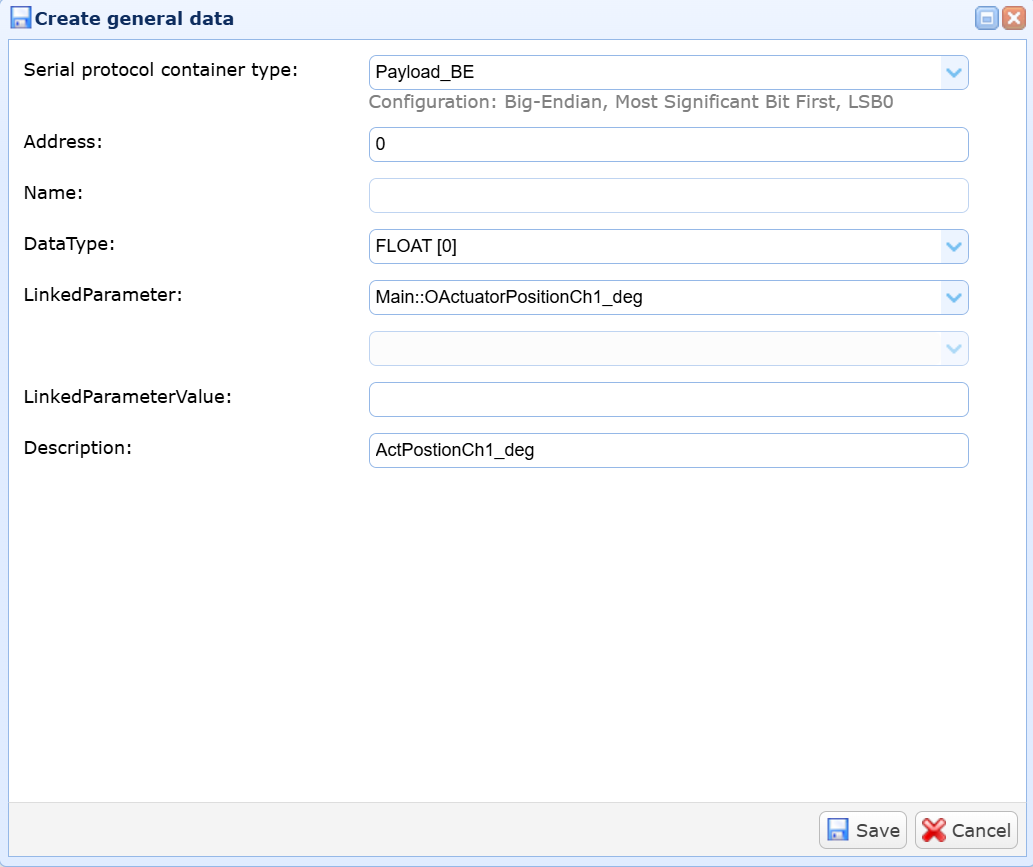}
\caption{Payload container creation form}
\label{fig:63a}
\end{figure}

Figure~\ref{fig:63a} shows the container creation form used to define a new payload element. As in the message-level workflow, the user interacts only with a limited set of engineering-relevant properties exposed through the configured representation.

In this example, the user specifies the container type (Payload\_BE), payload-relative address, data type, and description. Importantly, the address is defined relative to the beginning of the payload area rather than the beginning of the message. This allows engineers to work within a coordinate system consistent with protocol documentation and common engineering practice.

Low-level structural properties such as absolute addressing, encoding semantics, and container placement constraints are not specified manually at this stage. Instead, they are derived automatically from the definitions made earlier.

After creation, the new container becomes part of the complete structural representation of the message.

\begin{figure}[!htbp]
\centering
\includegraphics[width=.8\textwidth]{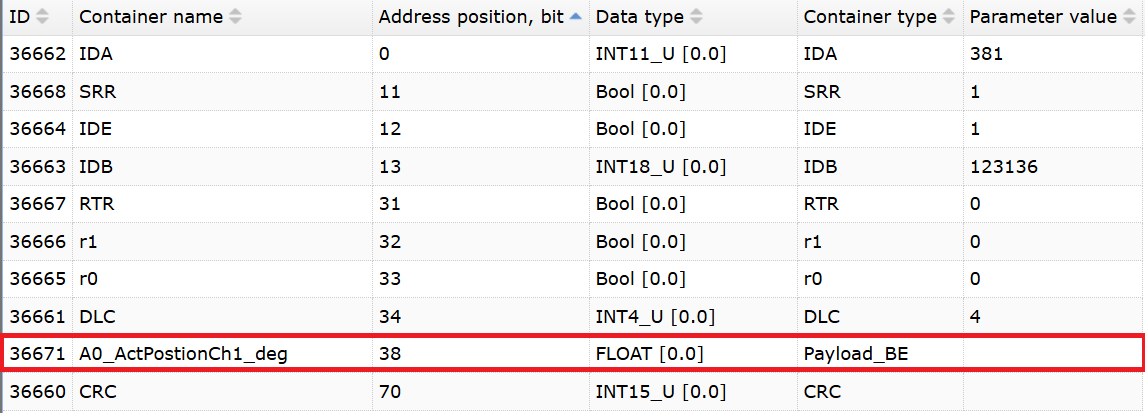}
\caption{Raw structural representation after container creation}
\label{fig:63b}
\end{figure}

Figure~\ref{fig:63b} shows the raw structural definition following container insertion. The newly created payload container appears within the unified bit-addressed message structure at the correct absolute offset. In this example, the container is placed at bit position 38, reflecting automatic conversion from payload-relative addressing to the absolute coordinate system used by the formal model.

The same container is also visible in the user-oriented representation.

\begin{figure}[!htbp]
\centering
\includegraphics[width=.8\textwidth]{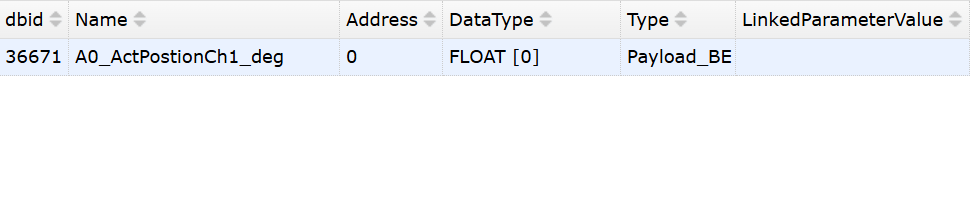}
\caption{User representation of container}
\label{fig:63c}
\end{figure}

Figure~\ref{fig:63c} shows the configured payload-level representation of the same container. In this view, the container is presented using payload-relative addressing and engineering-oriented attributes such as name, data type, and linked parameter. Containers of roles other than "payload" are hidden, as they are not required for routine engineering tasks.

Together, these examples illustrate that container definition follows the same principles established earlier for message-level operations: engineers work with domain-relevant abstractions, while the tool maintains a fully explicit structural representation underneath.

\subsection{ARINC 825 Message Representation Example}

As noted above, ARINC 825 is based on the CAN EFF frame format. Consequently, the message created in Section 6.2 can also be presented using an ARINC 825-oriented representation.

\begin{figure}[!htbp]
\centering
\includegraphics[width=.8\textwidth]{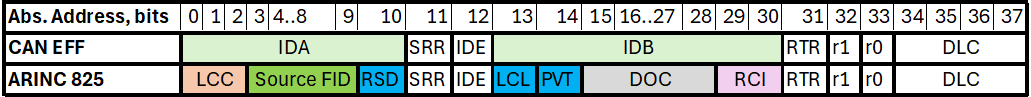}
\caption{Comparison of CAN EFF and ARINC 825 header interpretations}
\label{fig:64a}
\end{figure}

Before demonstrating the configured representation, it is useful to recall the main difference between the conventional CAN EFF view and the ARINC 825 interpretation of the same frame header. In common CAN engineering practice, the identifier is usually handled as a single decimal or hexadecimal value. In ARINC 825, however, the same header is interpreted as a composition of several subfields:

\begin{itemize}
\item LCC - Logical Communications Channel,
\item Source FID - Source Function Code Identifier,
\item RSD - Reserved bit,
\item LCL - Local bit,
\item PVT - Private bit,
\item DOC - Data Object Code,
\item RCI - Redundancy Channel Identifier.
\end{itemize}
Figure~\ref{fig:64a} illustrates the conceptual correspondence between the CAN EFF header and its ARINC 825 interpretation.

Using the configurable representation mechanism, the frame F145358852\_ActPosition, introduced in Section 6.2, can be presented in ARINC 825 form without modifying its underlying structural definition.

\begin{figure}[!htbp]
\centering
\includegraphics[width=.8\textwidth]{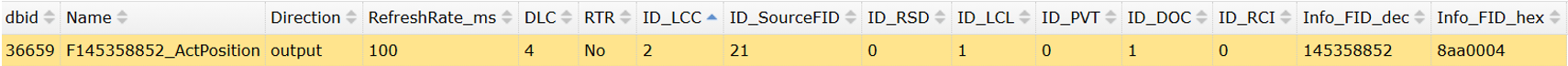}
\caption{ARINC 825-configured representation of CAN EFF frame}
\label{fig:64b}
\end{figure}

In the representation presented at Fig.~\ref{fig:64b}, the identifier is no longer shown as a single aggregate numeric value. Instead, it is decomposed into ARINC 825-specific subfields, which provides a more meaningful engineering interpretation of the same underlying header structure:

\begin{itemize}
\item LCC = 2 --- Normal Operation Channel
\item Source FID = 21 --- Source device identifier within the network
\item RSD = 0
\item LCL = 1 --- Message is local
\item PVT = 0 --- Message is not private
\item DOC = 1 --- Identifier of the data object within the source device message set
\item RCI = 0 --- Redundancy channel identifier is either zero or not used
\end{itemize}
The decimal and hexadecimal header values are retained in the representation for reference. This example demonstrates that the proposed method allows the same formal message model to support multiple engineering views, depending on domain-specific interpretation requirements, without changing the underlying frame structure.

\subsection{Ethernet-IP-UDP Message Representation Example}

The UDP-based example used in this paper contains the most structurally extensive header among the considered message types. For this reason, it provides a good use-case demonstration of how the proposed modeling approach handles message definitions with a large number of automatically maintained structural elements. The purpose of this example is to show that even a highly detailed message model can be created and edited in practice without overloading the user with routine low-level input.

\begin{figure}[!htbp]
\centering
\includegraphics[width=.8\textwidth]{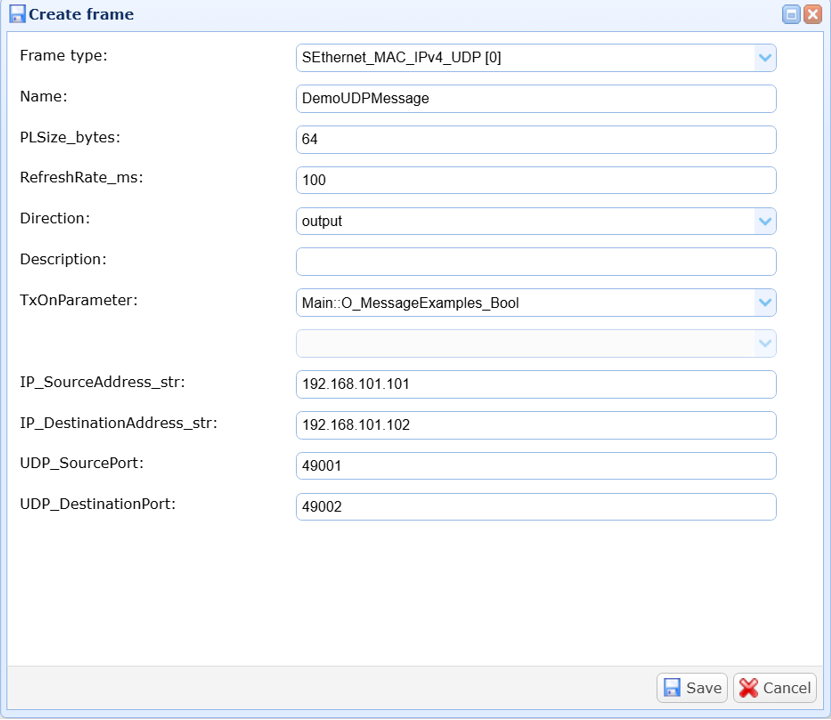}
\caption{"Create Frame" form for UDP}
\label{fig:65a}
\end{figure}

The first step is to create the message through the configured user-facing form as shown in Fig.~\ref{fig:65a}.

\begin{figure}[!htbp]
\centering
\includegraphics[width=.8\textwidth]{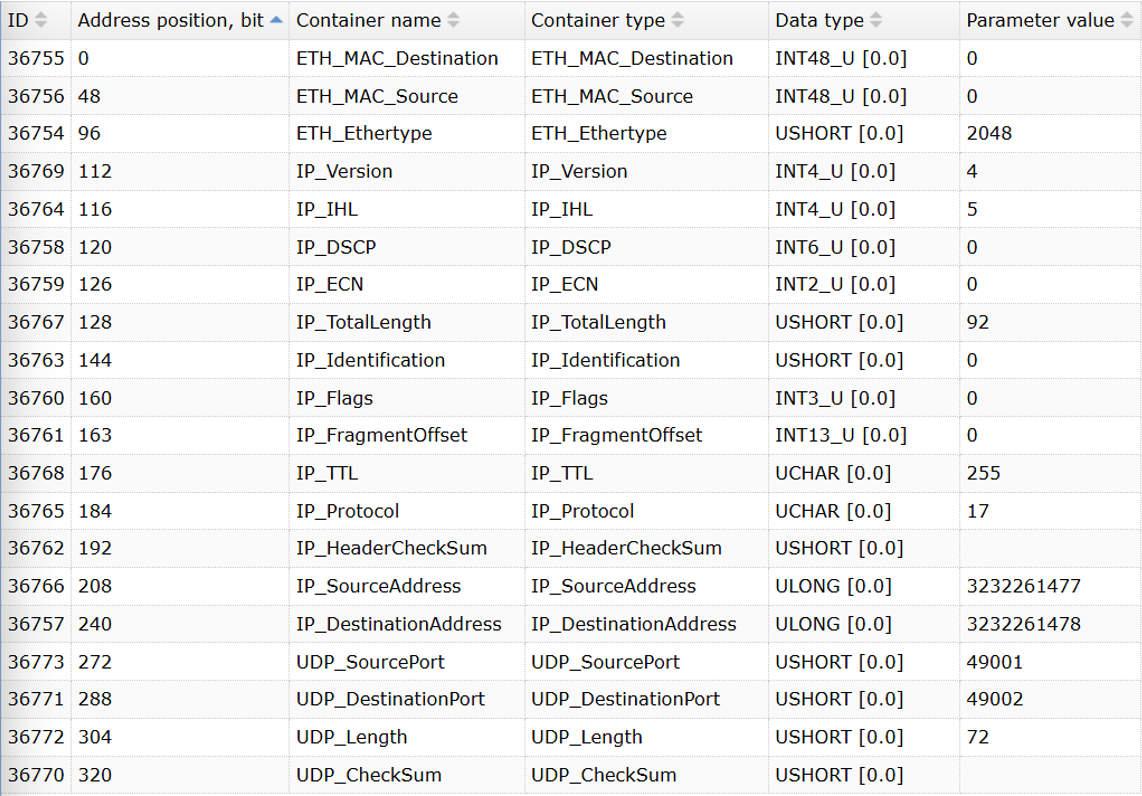}
\caption{UDP message raw structural representation after creation}
\label{fig:65b}
\end{figure}

After creation, the tool automatically generates the mandatory message structure shown in Figure~\ref{fig:65b}.

This example illustrates several important aspects of the implemented approach.

First, although the complete Ethernet--IP--UDP header structure is present in the underlying model, only the subset relevant to the target engineering workflow needs to be exposed in the user-facing representation. In the configuration used for this example, MAC addresses are not required by the intended application and therefore remain unspecified. This does not limit the model itself and can be changed later by adapting the representation if such information becomes relevant.

Second, some values are represented differently in the raw message model and in the engineering-oriented view. For example, IP addresses are stored internally as machine-processable integer values, while the configured representation exposes them in the conventional "192.168.101.101" notation familiar to engineers.

Third, structural elements that can be derived automatically are maintained automatically. For example, length fields corresponding to different protocol layers are maintained automatically rather than entered manually by the user.

\begin{figure}[!htbp]
\centering
\includegraphics[width=.8\textwidth]{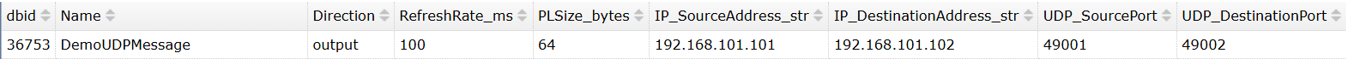}
\caption{UDP message user representation}
\label{fig:65c}
\end{figure}

Because the fully explicit header structure is not convenient for routine user interaction in this case, the configured engineering representation is limited to the properties typically needed in practice, such as IP addresses and UDP ports. An example of this user-oriented view is shown in Figure~\ref{fig:65c}.

\subsection{Model Consistency Enforcement}

In addition to providing structured mechanisms for message and container definition, the implemented environment includes automated consistency enforcement functions designed to ensure correctness of the resulting models.

These mechanisms operate continuously during model creation and modification and address several classes of structural and semantic constraints.

First, the tool verifies structural consistency of container placement within the message. In particular, it detects overlapping container definitions, ensures that all containers remain within message boundaries, and validates placement against constraints defined by the selected message type.

Second, the tool enforces constraints on permissible data types. Container definitions are validated against message-type rules governing allowable data representations, preventing structurally valid but semantically incorrect configurations.

Third, user-defined values are validated against configurable constraints designed to ensure downstream usability. For example, naming rules restrict the use of special characters that may be incompatible with target programming languages or code-generation environments.

Additional consistency checks address a broader range of project-specific requirements, including constraints derived from coding standards, integration rules, and toolchain assumptions. A detailed discussion of these mechanisms lies beyond the scope of this paper; however, their presence is essential for practical deployment of the proposed modeling approach in large-scale engineering workflows.

Together, these consistency enforcement mechanisms complement the structural definition capabilities of the model and play a key role in enabling reliable automation of transport-layer software generation and interface verification activities.
\clearpage

\subsubsection{Validation Against the Stated Requirements}

The proposed model, methods, and tool implementation were validated against the requirements formulated in Sections~4 and~5.

Table~\ref{tab:validation_traceability} summarizes this validation effort. It shows how the stated requirements are addressed by the proposed solution and where their fulfillment is demonstrated in the paper.

\begin{table}[!htbp]
\caption{Traceability between stated requirements and proposed solution}
\label{tab:validation_traceability}
\centering
\scriptsize
\setlength{\tabcolsep}{2.5pt}
\renewcommand{\arraystretch}{1.08}

\begin{tabularx}{\textwidth}{|
>{\raggedright\arraybackslash}p{0.06\textwidth}|
>{\raggedright\arraybackslash}X|
>{\raggedright\arraybackslash}p{0.2\textwidth}|
>{\raggedright\arraybackslash}p{0.2\textwidth}|}
\hline
\rowcolor{gray!20}
\textbf{Req.} & \textbf{Requirement text} & \textbf{Addressed by} & \textbf{Demonstrated in} \\
\hline

R1
& Protocol universality - The model must be capable of representing message formats used across heterogeneous
communication protocols, including both standardized and custom serial
message structures.
& Models of Data Type (4.2.2), Container (4.2.3), Message (4.2.4)
& Examples for ARINC 429/825, CAN EFF/BFF, MAVLink, and UDP/IP in Sections~4 and~6 \\
\hline

R2
& Complete message specification - A complete message instance must be constructible solely from information
contained in the model.
&  Models of Data Type (4.2.2), Container (4.2.3), Message (4.2.4)
& Structural examples and explicit message tables in Section~4 \\
\hline

R3
& Data representation flexibility - The model must not impose artificial or standard-driven restrictions on data types,
addressing schemes, or value representations used within message structures
& Addressing of all encoding details in models of data type (4.2.2) and Container (4.2.3)
& Structural examples with different encoding schemes in Section~4\\
\hline

T1
& Constraint-based consistency - The methods shall support maintaining consistency between message definitions and user-defined structural and semantic constraints
& Message type concept and related methods described in Section~5.3
& Message creation and editing presented in Sections 6.2 and 6.3  \\
\hline

T2
& Support for deterministic automation - The methods shall support automatic creation and completion of all message elements and properties that can be derived unambiguously from existing model data, constraint definitions, or user-provided engineering-level properties.
& Message type concept and related methods described in Section~5.3
& Message creation and editing presented in Sections 6.2 and 6.3  \\
\hline

T3
& User-oriented model interaction - The methods shall support presentation and editing of the formal model in a form convenient and familiar to the user.
& User representation method described in Section 5.4
& User representations given in Section~6 \\
\hline

T4
& User-configurable representations - The representation format used for model presentation and editing shall be configurable by the user and shall not be hard-coded.
& User representation method described in Section 5.4
& User representations given in Section~6  \\
\hline

\end{tabularx}
\end{table}

\clearpage
\section{Conclusion}

Efficient development of modern embedded systems increasingly depends on the use of automation toolchains. To be effective, such toolchains must rely on a solid formal basis. One of the key elements of this basis is a machine-readable ICD. To make such an ICD both practically useful and broadly applicable across heterogeneous communication environments, it should be built on a unified serial message model of the transport layer.

In this work, we proposed a unified and protocol-agnostic message model that makes it possible to describe serial messages explicitly and unambiguously, regardless of the protocol family to which they belong. The model is based on three elements: an explicit description of data types as the basic encoding units, a description of atomic message elements (containers) together with their required properties, and a description of complete message structure built from these atomic elements.

In addition to the model itself, the paper proposed methods for working with it in practice. These methods are based on two main mechanisms. The first is a configurable formal message type model, which supports both automation of routine operations and definition of structural constraints. The second is a configurable representation mechanism that allows the unified formal model to be presented and edited in forms convenient for engineering reading and editing.

The paper also demonstrated successful implementation of the proposed model and methods in the industrial tool dBricks. Taken together, the model, the methods, and the tool implementation form a workable foundation for automation toolchains intended to support machine-readable ICD development, automated generation of transport-layer software and all related transport-layer engineering tasks.

\bibliography{sn-bibliography}


\end{document}